\documentclass[twocolumn]{aastex63}

\usepackage{bm}
\usepackage{amsmath}

\usepackage{graphicx}

\usepackage{float}
\shorttitle{Multi-wavelength Radiation Properties in Pulsar Dissipative Magnetospheres}
\shortauthors{Yang \& Cao}
\begin{document}

\title{Modeling the Multi-wavelength Radiation Properties in Pulsar Dissipative Magnetospheres}

\author{Xiongbang Yang}
\affiliation{Department of  Mathematics, Yunnan University of Finance and Economics, Kunming 650221, Yunnan, P. R. China;gcao@ynufe.edu.cn}
\author{Gang Cao}
\affiliation{Department of  Mathematics, Yunnan University of Finance and Economics, Kunming 650221, Yunnan, P. R. China;gcao@ynufe.edu.cn}

\begin{abstract}
   We explore the multiwavelength radiation properties of the light curves and energy spectra in the dissipative magnetospheres of pulsars. The dissipative magnetospheres are simulated by the pseudo-spectral method with the combined force-free and Aristotelian electrodynamics, which can produce self-consistent accelerating electric fields mainly distributed in the equatorial current sheet outside the light cylinder. The multiwavelength light curves and spectra are computed by using the multiple emission mechanisms of both the primary particles accelerated by the accelerating electric fields in the equatorial current sheet and the secondary pairs with an assumed distribution spectrum. We then compare the predicted multiwavelength light curves and spectra with the observed data from the Crab, Vela, and Geminga pulsars. Our modeling results can systematically well reproduce the observed trends of the multiwavelength light curves and the spectra for these three pulsars.

\end{abstract}

\section{Introduction}\label{sect-intro}
The extreme circumstances of rapidly rotating and highly magnetized pulsars make it possible for them to be ideal accelerators in the Universe. They can emit very stable and regular electromagnetic radiation from accelerated high-energy particles in the radio up to the $\gamma$-ray band. In the past 15 years, the Fermi Large Area Telescope has provided us with abundant high-quality $\gamma$-ray light curves and spectra in the 100 MeV$-$300 GeV range. The Fermi $\gamma$-ray light curves generally show double-peak profiles and energy-dependent evolutionary patterns. The Fermi $\gamma$-ray spectra usually follow an exponential power-law shape with a cutoff energy around a few GeV \citep{abd13}. Recently, ground-based air-Cherenkov telescopes, such as MAGIC \citep{ali08,ale11,ale12,an16,acc20}, VERITAS, and H.E.S.S.\citep{abda18}, have detected sub-TeV $\gamma$-ray emission from a few $\gamma$-ray pulsars. The observed TeV spectra can be smoothly joined to those observed by Fermi, and perhaps indicate a transition from curvature radiation (CR) to the inverse-Compton scattering (ICS) regime. It is well known that the high-energy (HE) radiation from a pulsar is directly linked with the structure of its magnetosphere. Therefore, it is necessary to obtain a realistic structure of the magnetosphere to constrain the HE emission production sites and the emission mechanisms.

The analytic vacuum-retarded dipole derived by \citet{deu55} was widely used as the background field in the standard gap models to study pulsar spectra and light-curve properties. In these models, the CR \citep{rom96,hiro01,har08,tang08,pet19} or synchrotron curvature \citep{To18} from the primary particles was often thought to produce the GeV radiation, and the ICS radiation \citep{har08,ale11,pet11,du12,lyu12,lyu13,mo15,ru17} from the primary particles or the secondary pairs is considered as the origin of TeV photons. Actually, a realistic pulsar magnetosphere will deviate from the vacuum state and become a plasma-filled force-free (FF) one \citep{gol69}. The aligned FF magnetosphere was first obtained by \citet{con99} by solving the pulsar equation; they obtained an open$-$closed magnetospheric configuration with the equatorial current sheet outside the light cylinder (LC). The CKF solution was further confirmed by performing a time-dependent FF simulation \citep[e.g.,][]{ko06,mc06,tim06,par12,cao16a}. The three-dimensional inclined FF magnetosphere was first obtained by \citet{spi06} with the finite-difference time-domain method, which was further explored by \citet{kal09} with the inclusion of a perfectly matched layer, and by \citet{pet12} and \citet{cao16b} with the pseudo-spectral method. FF magnetospheres are also used to predict the radiation characteristics of pulsars \citep[e.g.,][]{bai10,con10,har15}.

Although the FF magnetosphere was thought to be closer to a realistic pulsar magnetosphere, it was inherently nondissipative, whereas realistic pulsar magnetospheres should have some dissipative regions that can self-consistently produce the observed pulsar emission. Therefore, the resistive model with a conductivity was developed to produce the dissipative regions. Resistive magnetospheres also became available \citep[e.g.,][]{kal12a,li12,cao16b} and were used to predict the pulsar $\gamma$-ray emission by the test particle trajectory method \citep{kal12b,kal14,kal17,bra15,cao19,yang21}. These studies revealed that the $\gamma$-ray radiation mainly comes from the outer magnetosphere near the equatorial current sheet outside the LC. Resistive magnetospheres can produce a self-consistent accelerating electric field by introducing a conductivity, but they cannot explain the microscopic physical origin of the conductivity itself. Therefore, the particle-in-cell (PIC) method was developed to model the pulsar magnetosphere by self-consistently treating the interactions of the particle motions, the emitting photons, and the electromagnetic fields. Some groups attribute the synchrotron radiation (SR, \citep{ce16,ph18}) from particles accelerated via magnetic reconnection in the current sheet to the main GeV emission mechanism, while other groups ascribe CR \citep{bra18,kal18,kal22,kal23} to the main GeV emission mechanism. However, the particle Lorentz factors obtained by the PIC method are much smaller than the real ones that can produce the observed GeV photons.

An alternative Aristotelian electrodynamics (AE), which can include the reaction of the emitting photons on the particle dynamics, has been proposed to model the pulsar magnetosphere
\citep{gru12,gru13}. Moreover, combined FF and AE magnetospheres are developed to simulate the pulsar magnetosphere, which can constrain the dissipative region only near the current sheet \citep{con16,cao20,pet20,pet22}. \citet{cao22} further calculate the CR energy spectra and light curves from the primary particles in the combined FF and AE magnetospheres by the particle trajectory method. Furthermore, the Fermi energy-dependent double-peak light curves, and the phase-averaged and phase-resolved GeV spectra, can also be reproduced for a high pair multiplicity, indicating that particle acceleration and $\gamma$-ray emission mainly originate in the current sheet outside the LC \citep{cao24}. However, the individual Fermi $\gamma$-ray band is not yet enough to distinguish between different emission locations and emission mechanisms in the magnetospheres. The study of combined multiwavelength light curves and spectra is thus expected to put a stronger constraint on these emission locations and mechanisms. Recently, \citet{har15} and \citet{har18,har21} explored the multiwavelength radiation properties of pulsars in the FF magnetosphere with a constant accelerating electric field distribution. However, they cannot well reproduce the peak phases of the observed multiwavelength light curves, indicating that a radially and azimuthally dependent accelerating electric field distribution is needed in modeling the pulsar multiwavelength radiation.

In the paper, we further expand the study of \citet{cao22,cao24} by simultaneously computing the multiwavelength radiation properties of pulsars with multiple radiation mechanisms from both the primary particles accelerated by the self-consistent accelerating electric field in the equatorial current sheet and the secondary pairs with an assumed pair spectrum in the combined FF and AE dissipative magnetospheres. The modeled multiwavelength light curves and spectra are then directly compared with those of the Crab, Vela, and Geminga pulsars. In Section 2, we describe the dissipative magnetospheric model and the multiwavelength radiation model. In Section 3, we apply our model to explain the multiwavelength light curves and spectra of the Crab, Vela, and Geminga pulsars. Finally, the conclusions and discussions are presented in Section 4.

\section{Model}\label{MultiRadiation}
\subsection{The combined FF and AE magnetospheres}

The pulsar magnetosphere was obtained by solving the time-dependent Maxwell equations in the comoving frame
\begin{eqnarray}
{\partial {\bf B}\over \partial t'} &=& -{\bf \nabla} \times ({\bf E}+{\bf V}_{\rm rot}\times{\bf B})\; ,\\
{\partial  {\bf E}\over \partial t'} &=& {\bf \nabla} \times ({\bf B}-{\bf V}_{\rm rot}\times{\bf E})-{\bf J}+{\bf V}_{\rm rot}\nabla\cdot{\bf E}\; ,\\
\nabla\cdot{\bf B} &=& 0\;,\\
\nabla\cdot{\bf E} &=& \rho_{\rm e}\;,
\end{eqnarray}
where $\bf B$ is the magnetic field, $\bf E$ is the electric field, $\rho$  is the charge density, $\bf J$ is the current density, and ${\bf V}_{\rm rot}$  is the corotating velocity. It is noted that all quantities are defined in the inertial observed frame(IOF).

The FF magnetospheres are thought to be closest to the realistic pulsar magnetosphere, but they cannot include any dissipation to produce the observed multiwavelength radiation. Recently, the combined FF and AE magnetospheres have been developed to approximate a realistic pulsar magnetosphere, which can indirectly include the reaction of the emitting photons on the particle dynamics and introduce the local dissipation where the FF condition is violated. In this paper, we use the combined FF and AE magnetospheres to approximate a realistic pulsar magnetosphere and predict the pulsar multiwavelength radiation. The current density {\bf J} in the combined FF and AE magnetospheres is defined by the local electromagnetic fields as \citep{cao20,cao22}
\begin{eqnarray}\label{eq-current}
{\bf J }=  \rho_e  \frac {{\bf E} \times {\bf B}}{B^2+E^2_{0}}+ (1+\kappa)\left|\rho_e\right| \frac{ (B_0{\bf {B}}+E_0{\bf {E}}) }{ B^2+E^2_{0}}\;,
\end{eqnarray}
where $E_{0}$ and $B_{0}$ are the electric and magnetic fields in the fluid frame where ${\bf {E}}$ is parallel to ${\bf {B}}$. They are related by the following conditions: $B^2_{0}-E^2_{0}={\bf B}^2-{\bf E}^2$, $E_{0}B_{0}={\bf E}\cdot {\bf B}, E_{0}\geq0$, where $E_{0}$ is the accelerating electric field parallel to the magnetic field. $\kappa$ is a pair multiplicity phenomenologically connected with the pair cascade process, and is introduced to adjust the strength and the distribution of the accelerating electric field $E_0$ in the equatorial current sheet.

We simulate a series of the dissipative magnetospheres for the magnetic inclination angles $\chi$  from $0^\circ$ to $90^\circ$ in $5^\circ$ intervals, where the FF description is enforced in the $E<B$ regions and the AE description is implemented in the $E\geq B$ regions. The electric and magnetic fields are discretized in the regions from $r_{\star}=0.2{R_{\rm LC}}$ to $r_{\rm max}=3{R_{\rm LC}}$. The pair multiplicities are taken as $\kappa=\{0,1,3\}$. Moreover, a high resolution of $N_r\times N_\theta \times N_\phi = 129\times64\times128$ is used to better resolve the equatorial current sheet. Lastly, a nonreflecting boundary condition is implemented to avoid inward reflection from the outer boundary so as to ensure the combined FF and AE magnetospheres reach a stationary state for several rotation periods.

\begin{figure}[htb]
\centering
\begin{tabular}{c}
\hspace*{-0.8cm}
\includegraphics[width=8.5 cm,height=6 cm]{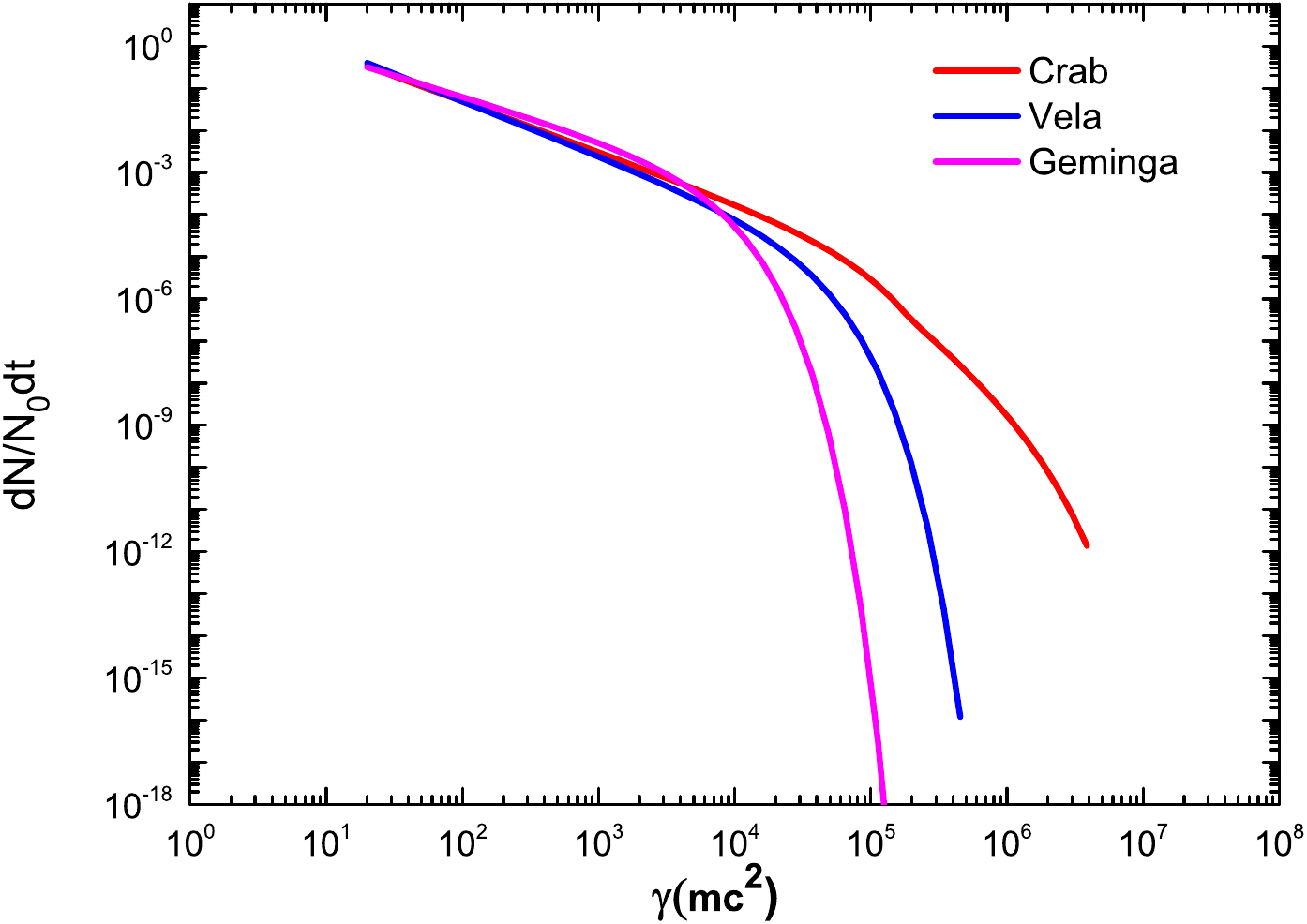}\,
\end{tabular}
\caption{We plot the assumed normalized pair spectra for Crab, Vela, and Geminga pulsars. The pair spectra of Vela and Geminga pulsars display an exponential power-law shape, while that of the Crab should be presented as a piecewise exponential power law.}
\label{pairs}
\end{figure}

\begin{figure}[htb]
\centering
\begin{tabular}{c}
\hspace*{-0.8cm}
\includegraphics[width=8.5 cm,height=6 cm]{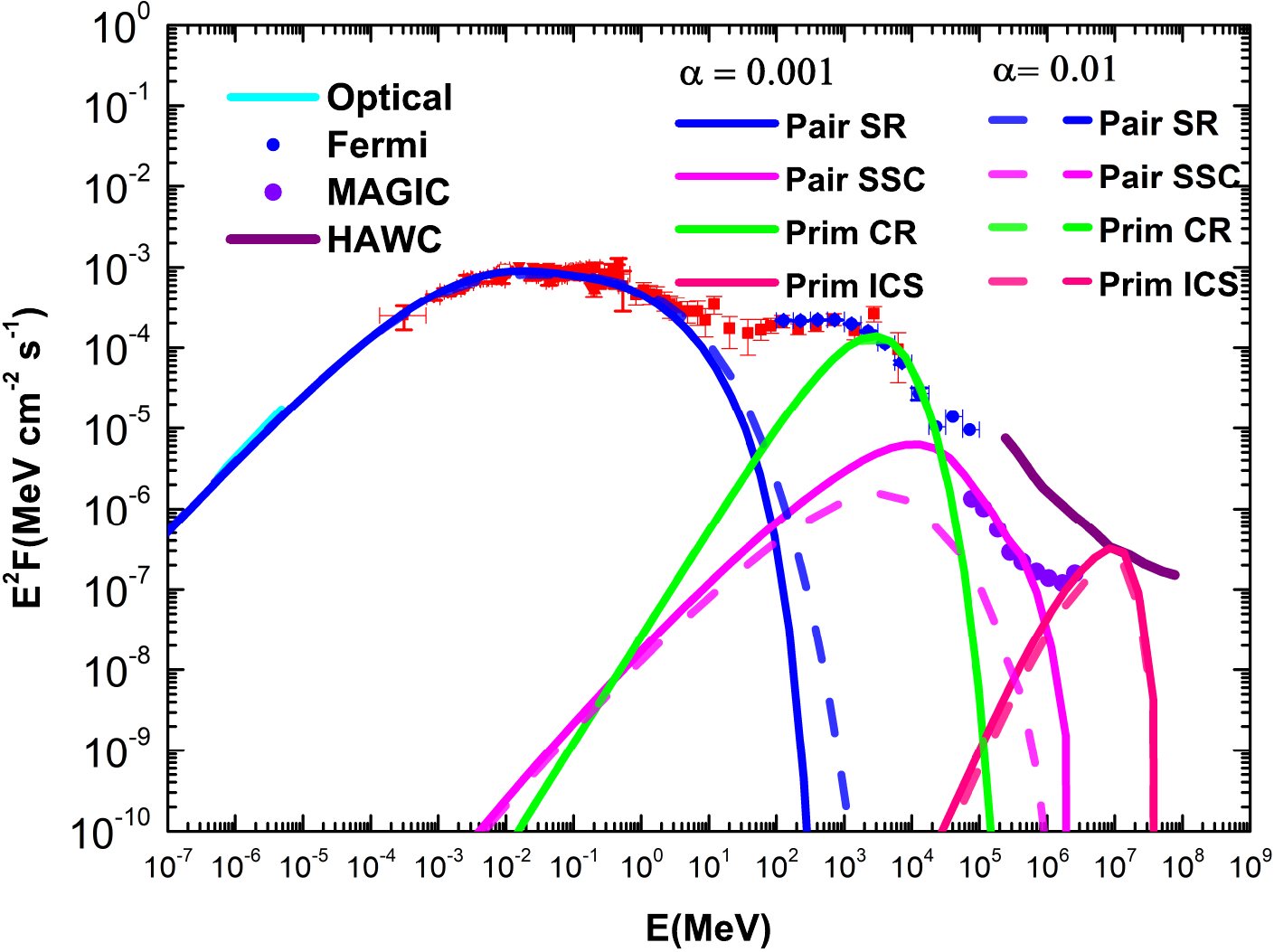}\,
\end{tabular}
\caption{The broadband SEDs of the Crab pulsar for the magnetic inclination $\chi=75^\circ$, viewing angle $\zeta=57^\circ$, and the pair multiple number $\kappa=3$ in the current density of equation (\ref{eq-current}). The pair multiplicity takes the value $\rm M_{pair}=3\times 10^5$ and the pitch angle $\rm \alpha=0.001$ is shown as the solid lines. For comparison, the predicted SEDs for a larger pitch angle $\rm \alpha=0.01$ are also shown as the dashed lines.  We can see that the lower $\alpha$ produces the well matched SEDs with the multiwavelength SED. The primary CR spectra for these two $\alpha$ are identical because they are independent of the pitch angle. See the context in the results for more details about the discussions of the effect of the pitch angle on the SEDs. Observed data are taken from \citet{kui01}, \citet{abd13}, \citet{an16}, and \citet{sol19}. The 50 hr HAWC observational sensitivity curve \citep{abe17} is also plotted.}
\label{CrabSED}
\end{figure}

\subsection{The injecting and tracking of the primary particles and the secondary pairs}\label{sect-PIT}
It is commonly recognized that the charged particles can be extracted from the pulsar PC surface and rapidly accelerated by the rotating induced electric field to extremely relativistic velocity to produce HE photons, which will be converted in bursts to the secondary pairs through the strong magnetic field in multiple pair cascade processes. This will cause the pulsar magnetosphere to reach the FF state in most regions. The secondary pairs can still produce low-energy SR photons, which may be boosted to HE photons through the ICS or synchrotron self-Compton (SSC) processes.

In our previous work, we only produce the  pulsar Fermi $\gamma$-ray emission by computing the CR radiation from the primary particles \citep{cao22,cao24}. In this paper, we will further produce the pulsar multiwavelength radiation by computing the SR, CR and ICS spectra from both the primary particles and the secondary pairs. We will inject a series of the primary  particles and the secondary pairs  from the PC surface in the specified open volume coordinates of $( \rm {r}_{ovc}, \rm {\phi}_{m} )$ \citep{dyk04}. The radial coordinate $\rm r_{ovc}=1$ denotes the PC rim, and $\rm r_{ovc}=0$ is the magnetic pole. The azimuthal coordinate $\rm {\phi}_{m}$ is the magnetic phase around the PC in the counterclockwise direction, which is divided into 360 equal portions. Primary particles with low initial Lorentz factor($\rm \gamma_{0}=100$) are uniformly injected from the PC surface in the range $\rm {r}_{ovc}=0.9-1$ with 0.01 intervals, and will be accelerated by the self-consistent accelerating electric field along their trajectories. Secondary pairs are injected with an assumed pair spectrum in the range $\rm {r}_{ovc}=0.8-0.9$ with 0.01 intervals, but they do not undergo any acceleration along their trajectories. The trajectories of both the primary particles and secondary pairs are tracked from the neutron surface up to $\rm 2.5R_{LC}$ in the IOF by
\begin{eqnarray}\label{eq-velocity}
\frac{d{\bf x}}{dt} &=& \bf v_{\pm}\; ,\\
\bf v_{\pm} &=& { {\bf E} \times {\bf B}\pm(B_0{\bf {B}}+E_0{\bf {E}}) \over B^2+E^2_{0}}\; .
\end{eqnarray}
where the two signs correspond to positrons and electrons, which follow different trajectories in the combined FF and AE  magnetospheres.

For the primary particles, taking into account the influence of the local accelerating electric field and the radiation losses, their Lorentz factors ($\gamma$) along the trajectories are integrated by the following expression
\begin{eqnarray}\label{eq-gamma}
\frac{d\gamma}{dt}=\frac{q_{\rm e}c E_0}{m_{\rm e}c^2}- \frac{2q^2_{\rm e} \gamma^4}{3R^2_{\rm CR}m_{\rm e}c}.
\end{eqnarray}
Where $\rm q_e$ and $m_e$ are the charge of the electron and its rest mass, respectively. c is the velocity of light in free space, and $\rm R_{CR}=dl/{d\theta}$ is the curvature radius of the particle trajectories. $dl$ is the segment length within time dt and $\rm d\theta$ is the angle spanned by the adjacent velocity vectors along the particle trajectory. We compute the CR and ICS spectra for the primary particles along each trajectory. For the secondary pairs, the exact pair spectra need to be determined by performing a time-dependent pair cascade simulation. However, it is very difficult to perform a three-dimensional pair cascade simulation that includes time-dependent electromagnetic fields and particle production. Only the one-dimensional (1D) pair cascade simulation is available  in a time-dependent manner \citep{tim13,tim15}. Therefore, we approximate the secondary pair spectra as a power-law exponential cutoff shape with
\begin{eqnarray}
\rm N(\gamma) &\propto& \gamma^{-s} \exp(-\gamma/\gamma_{cut})\;,
\end{eqnarray}
where the $\gamma$ is the energy of secondary pairs, $\rm s$ is the spectral index, and $\rm \gamma_{cut}$ is the cutoff energy of the secondary  pairs. The 1D pair cascade simulation found that the pair cascade operates only in the supercurrent, $\rm J/{J_{GJ}}>1$, and anticurrent,$\rm J/{J_{GJ}}<0$, regions. Therefore, in addition to injecting the secondary pairs in the range of $\rm {r}_{ovc}$, we further limit the pair injection to the supercurrent ($\rm J/{J_{GJ}}>1$) and anticurrent ($\rm J/{J_{GJ}}<0$) regions. For the Crab pulsar, we assume the pair Lorentz factors ranging from $\rm \gamma_{min}=20$ to $\rm \gamma_{max}=5\times10^6$. The spectral index $\rm s_1=2.2$ and the cutoff lorentz factor $\rm \gamma_{cut1}=7\times10^4$ for $\rm \gamma_{min} \leq \gamma < 2\times 10^5$; $\rm s_2=3.6$ and  $\rm \gamma_{cut2}=8\times10^5$ for $\rm 2\times 10^5 < \gamma \leq \gamma_{max} $. For the Vela pulsar, the Lorentz factors range from $\rm \gamma_{min}=20$ to $\rm \gamma_{max}=6\times10^5$, $\rm s=2.3$ and $\rm \gamma_{cut}=1\times10^4$. For the Geminga pulsar, the Lorentz factors range from $\rm \gamma_{min}=20$ to $\rm \gamma_{max}=6\times10^5$, $\rm s=2.0$ and $\rm \gamma_{cut}=4\times10^3$. Their injected pair spectra adopted in the following calculations for the SR and SSC spectra of the secondary pairs are plotted in Figure \ref{pairs}.

\begin{figure*}[htp]
\centering
\begin{tabular}{c}
\includegraphics[width=4.2cm,height=4.1cm]{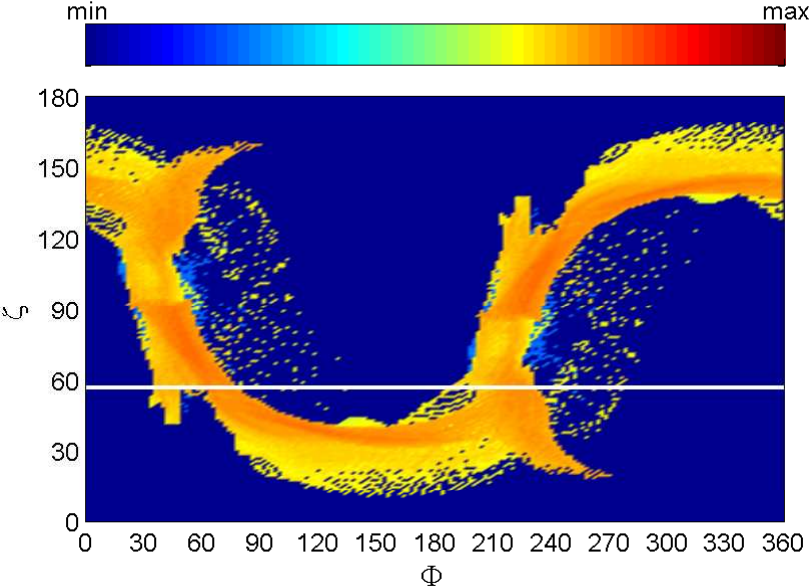}\,
\includegraphics[width=4.2cm,height=4cm]{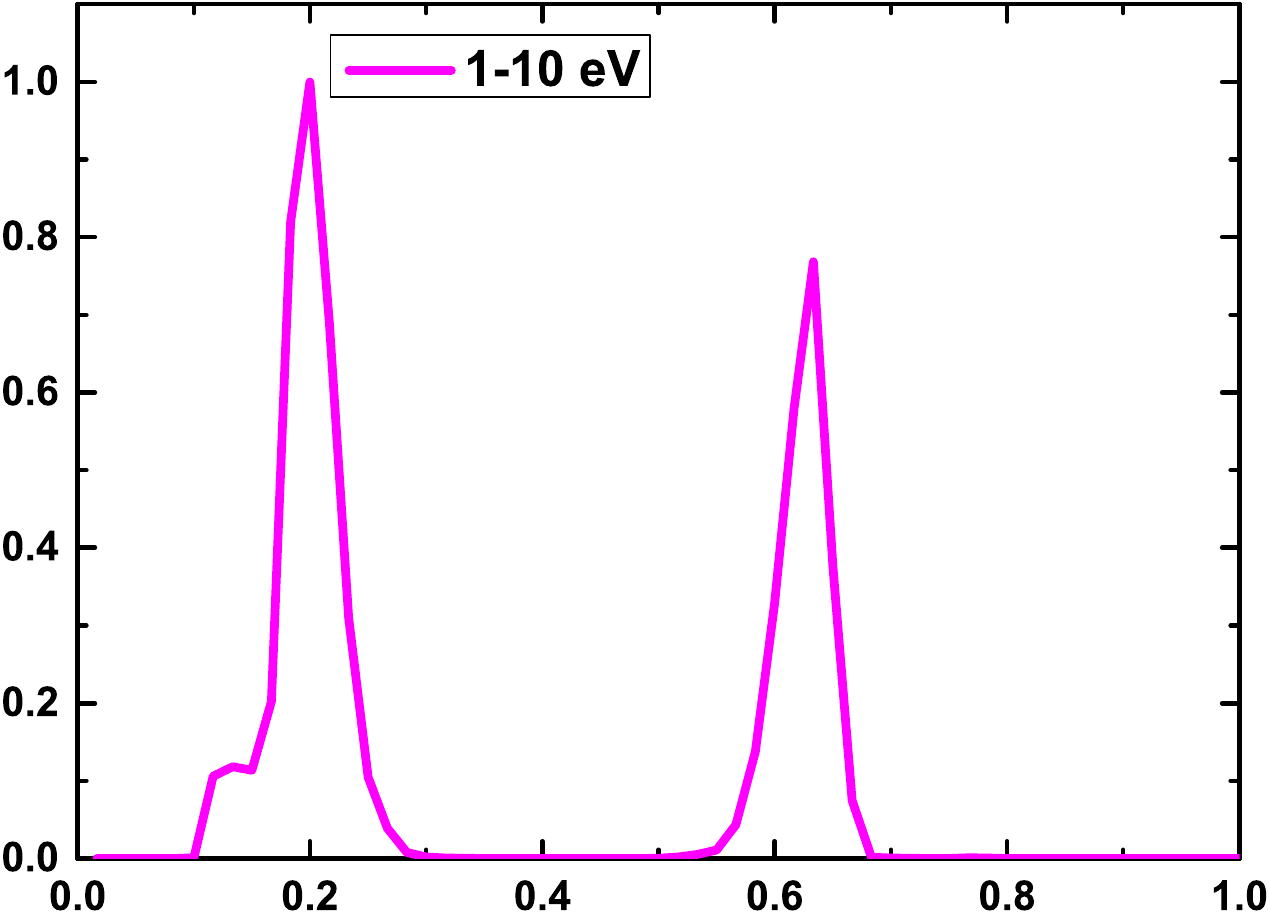}\,
\includegraphics[width=4.2cm,height=4cm]{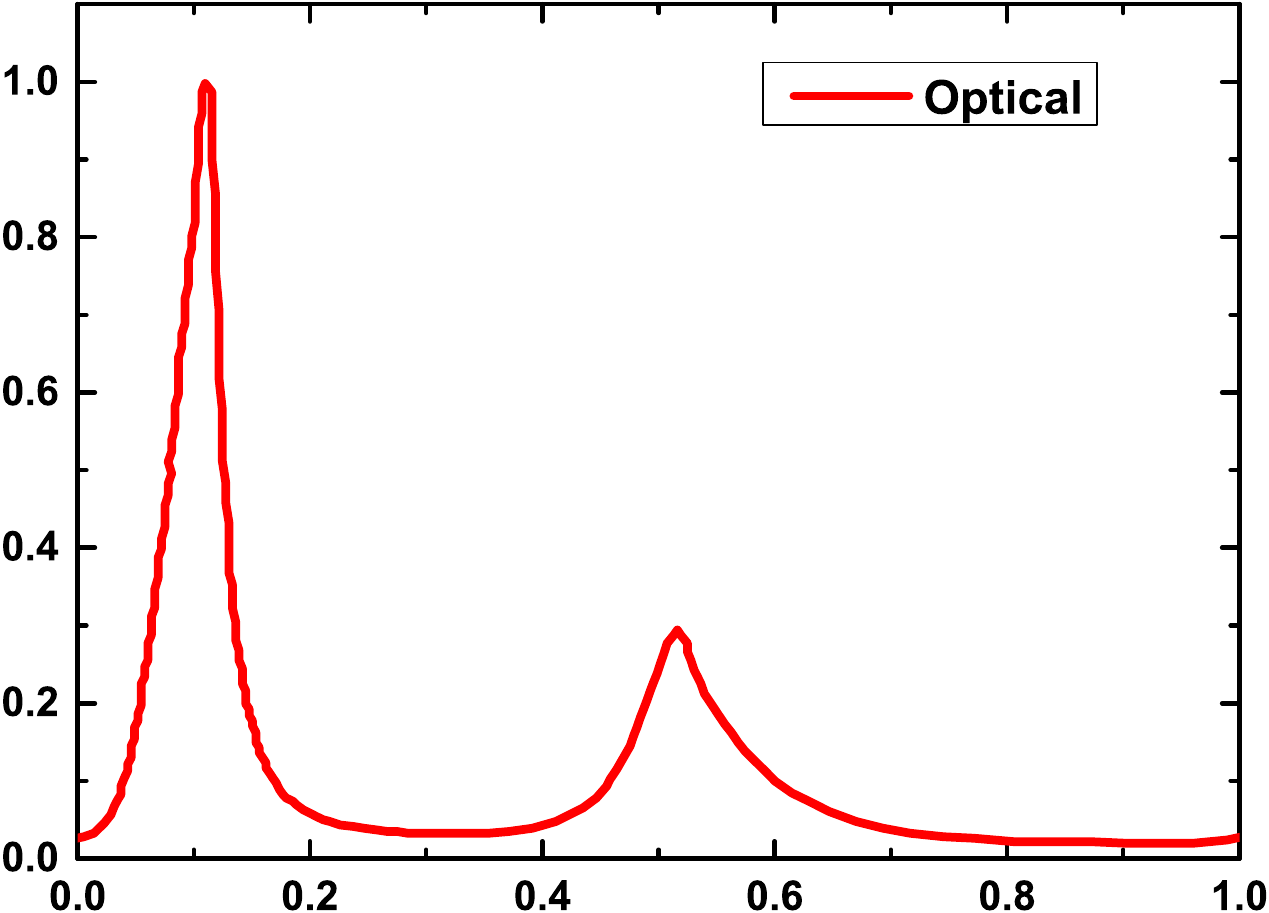}\\

\includegraphics[width=4.2cm,height=4.1cm]{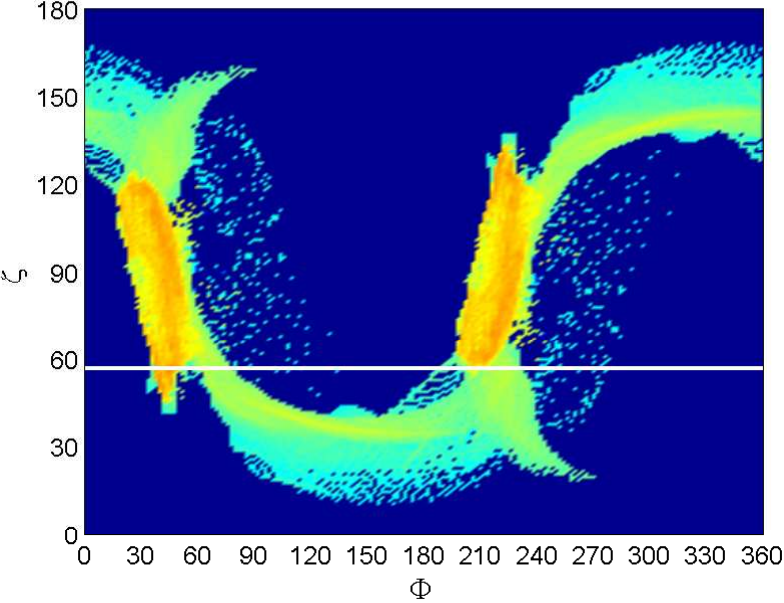}\,
\includegraphics[width=4.2cm,height=4cm]{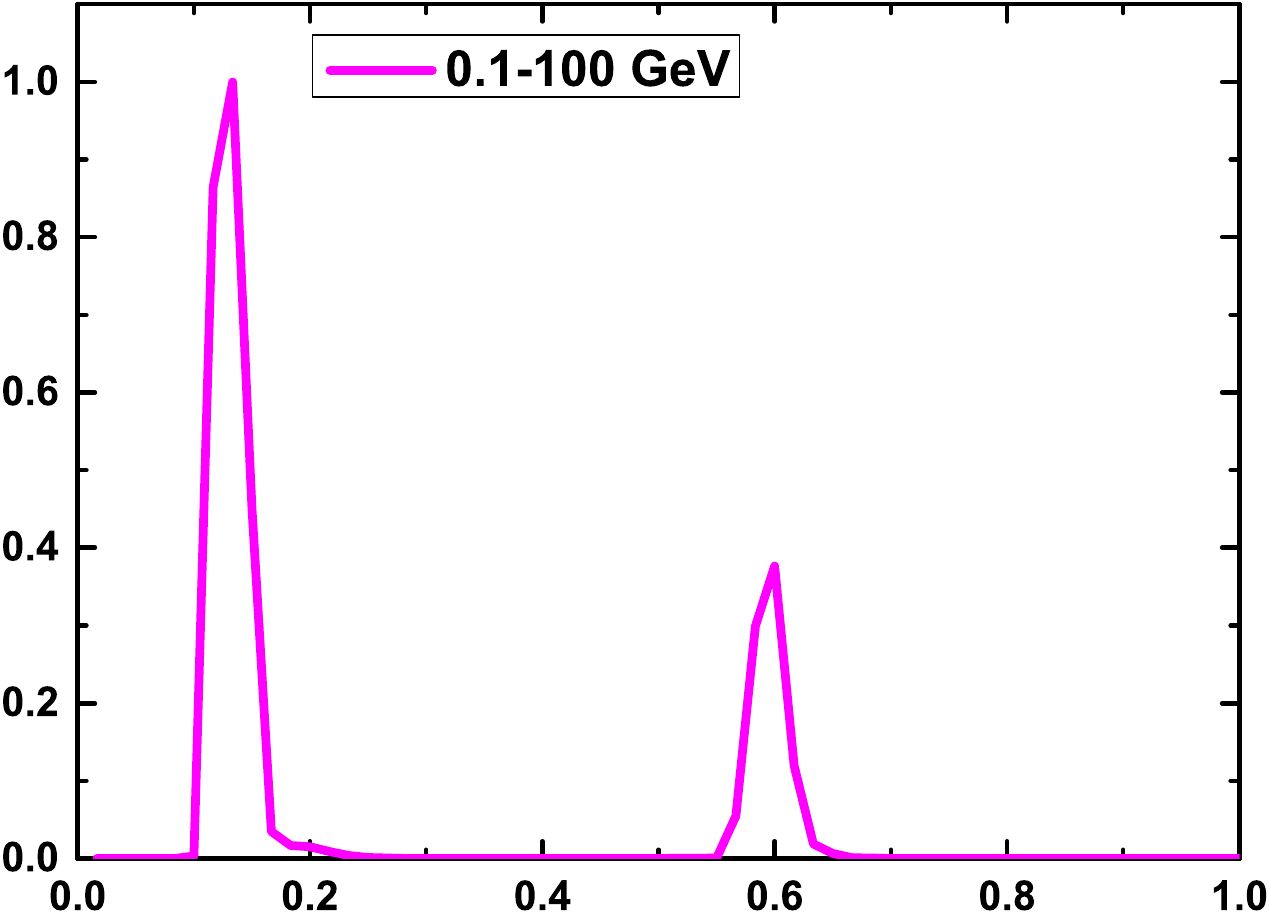}\,
\includegraphics[width=4.2cm,height=4cm]{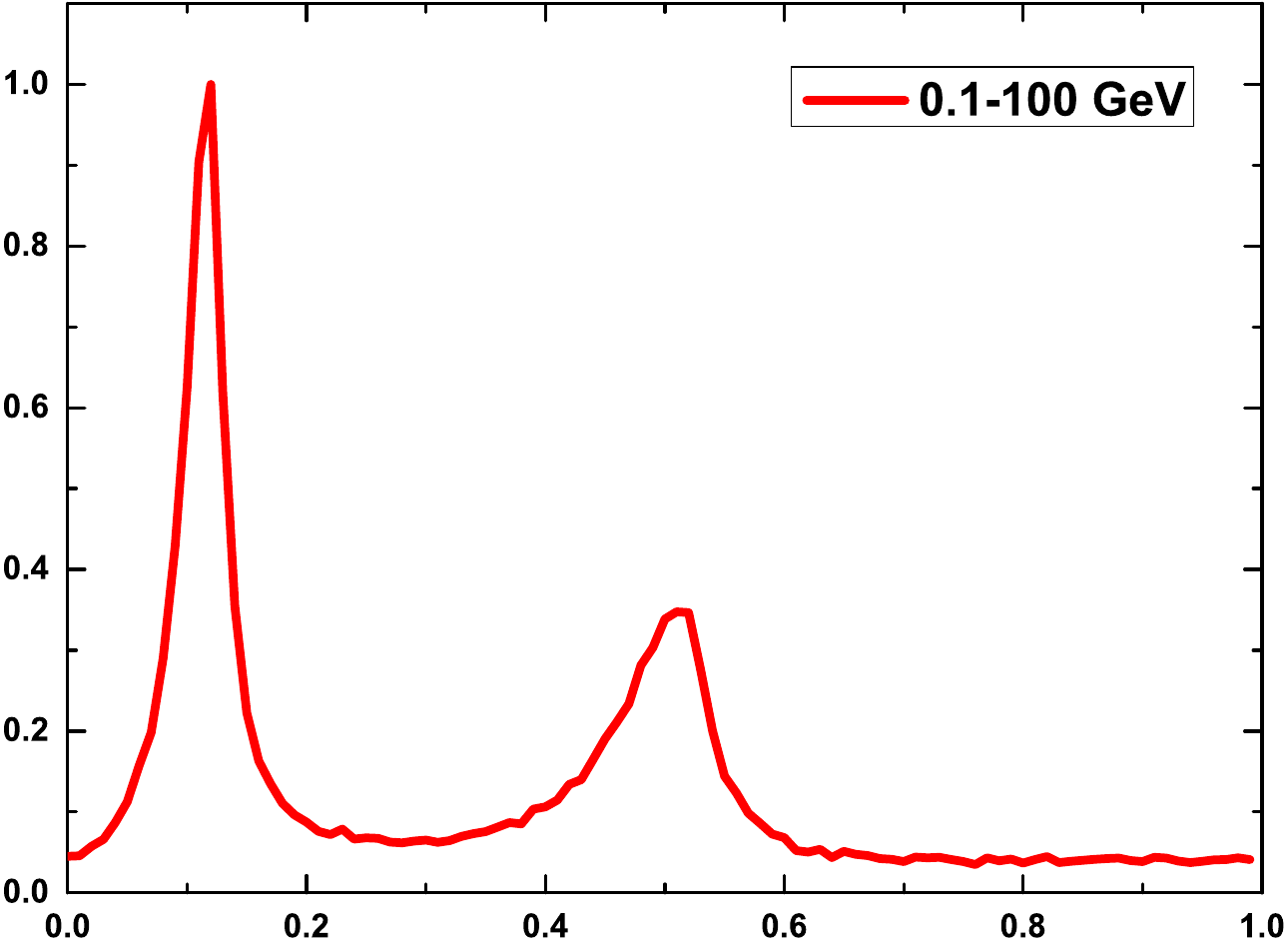}\\

\includegraphics[width=4.2cm,height=4.1cm]{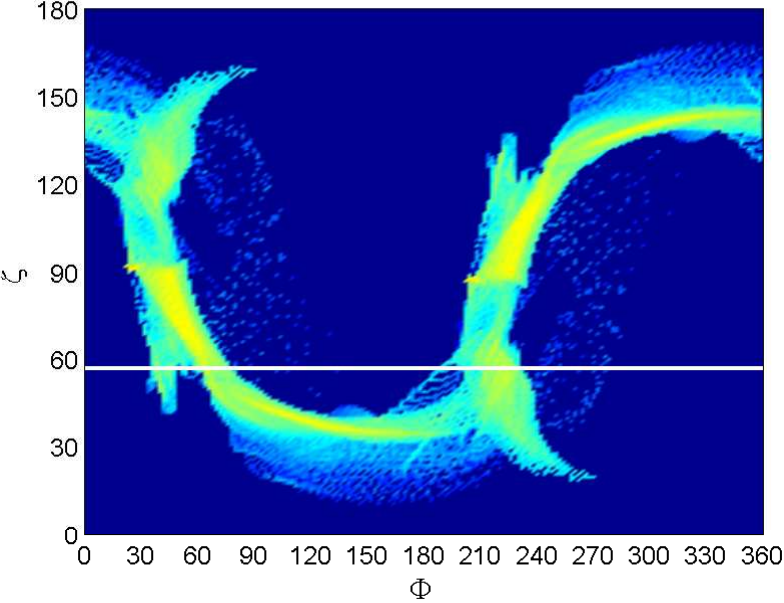}\,
\includegraphics[width=4.2cm,height=4cm]{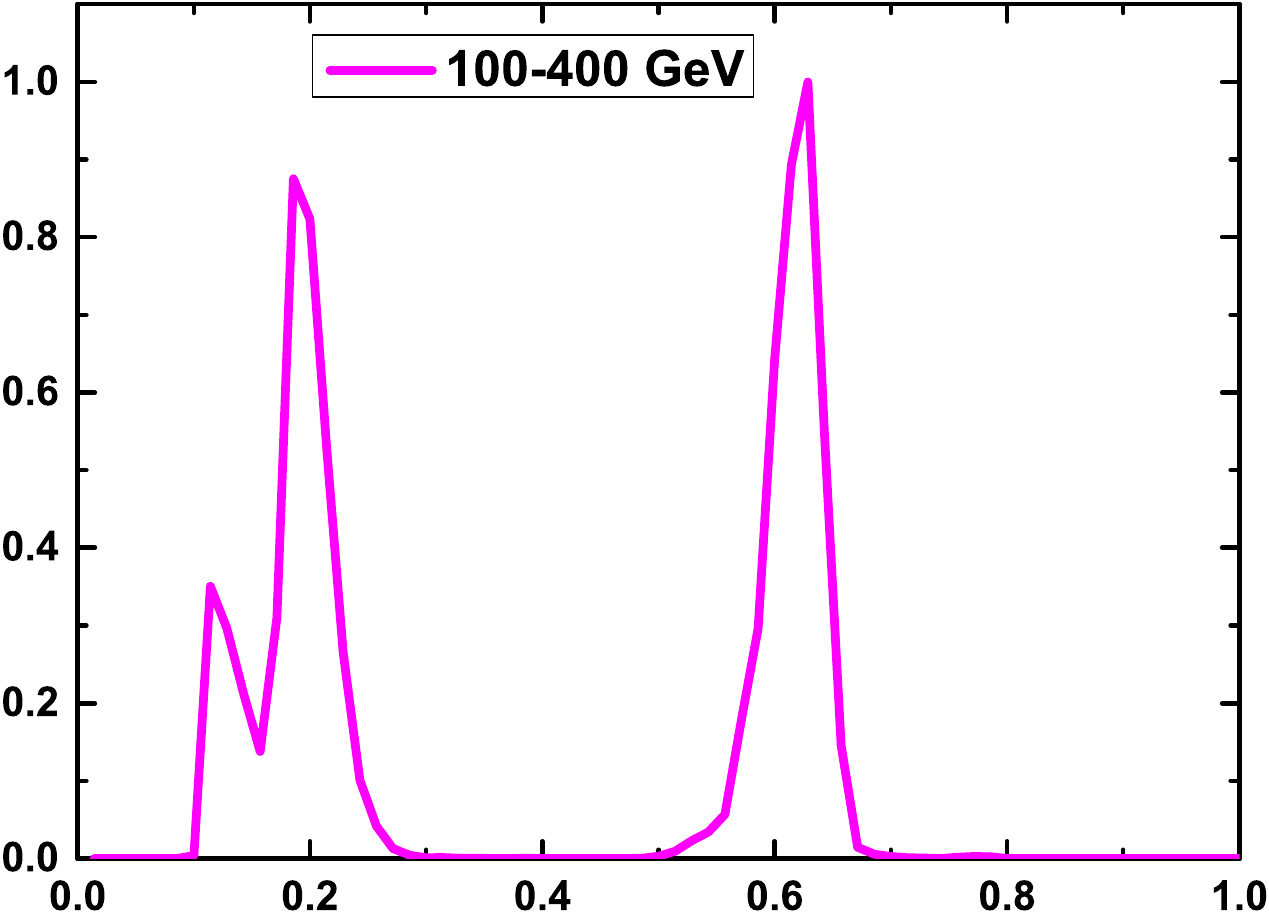}\,
\includegraphics[width=4.2cm,height=4cm]{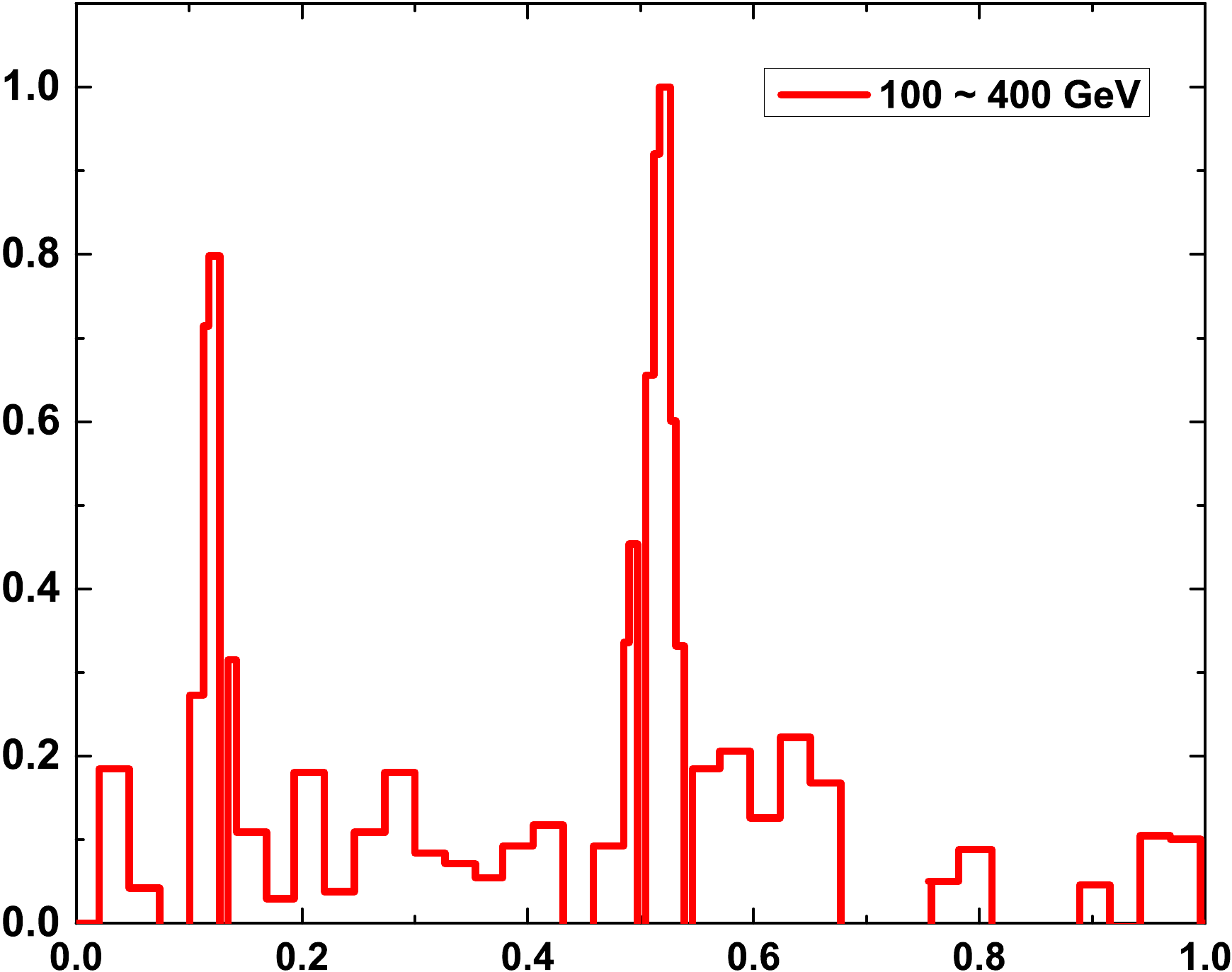}\\

\includegraphics[width=4.2cm,height=4.1cm]{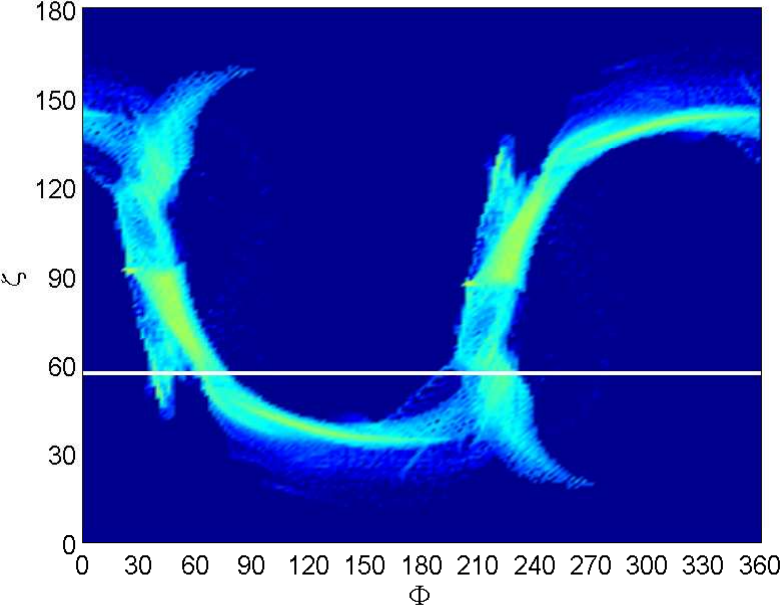}\,
\includegraphics[width=4.2cm,height=4cm]{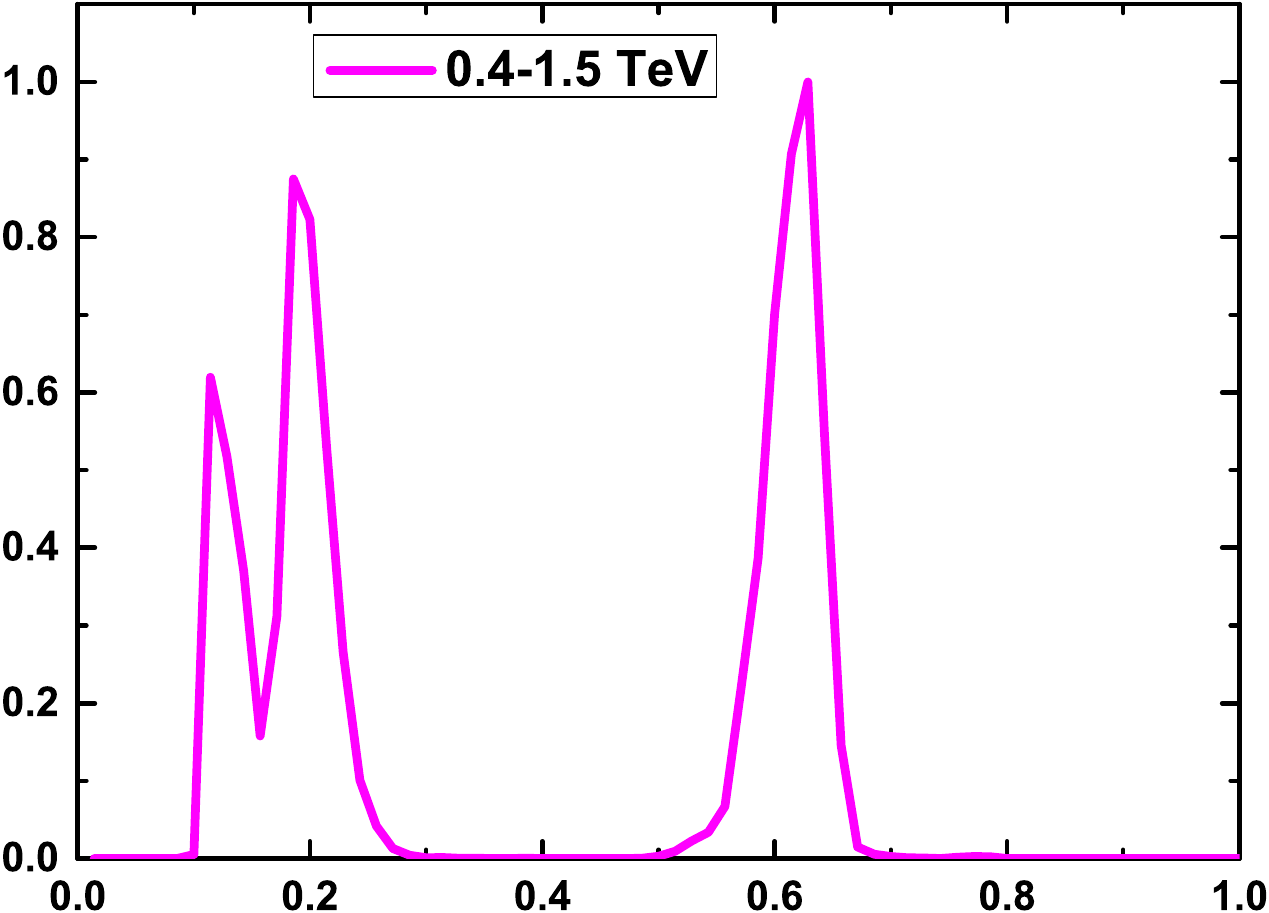}\,
\includegraphics[width=4.2cm,height=4cm]{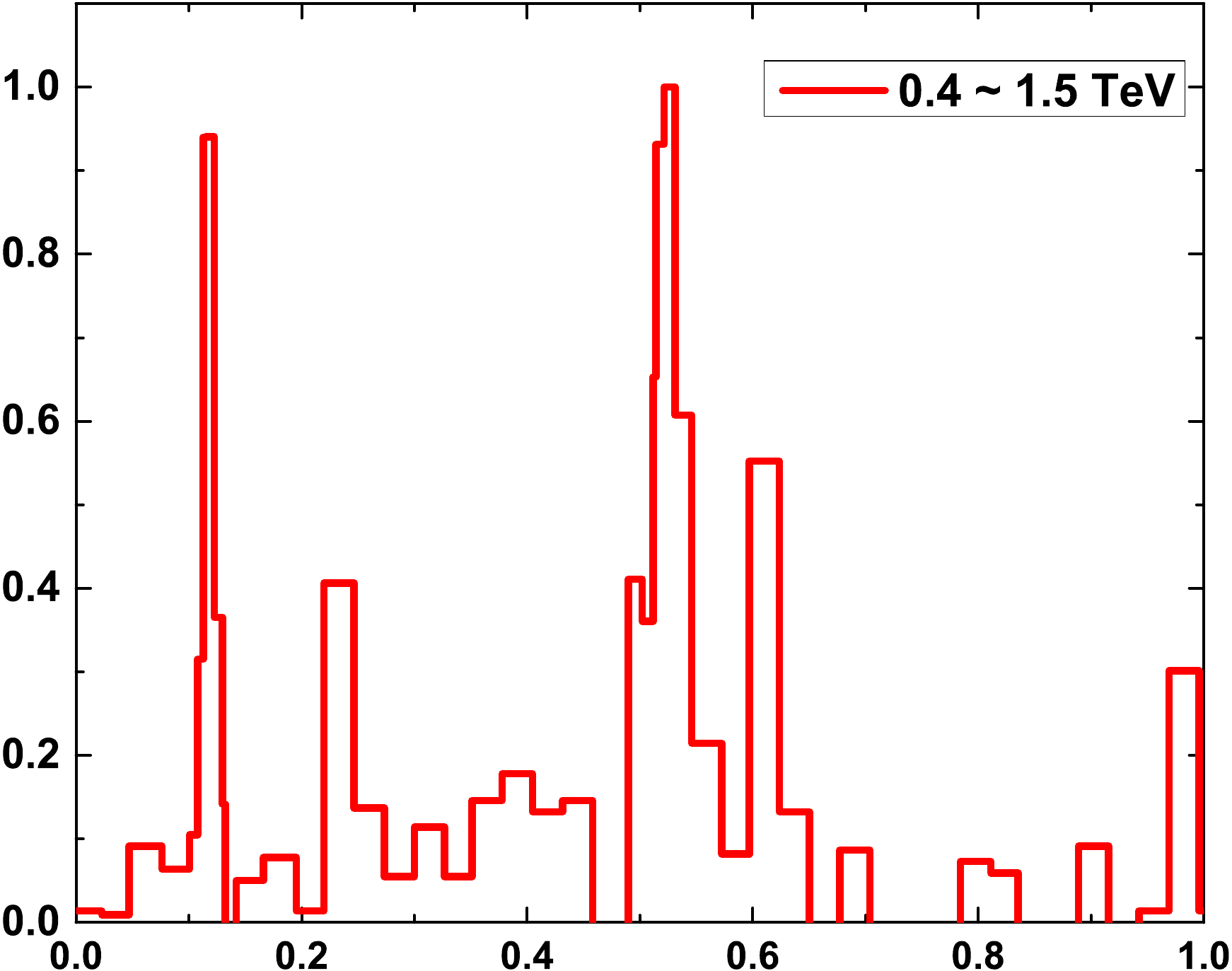}\\

\end{tabular}
\caption{The broadband sky maps and light curves for the Crab pulsar with the same parameters as in Figure \ref{CrabSED} for $\alpha=0.001$. Four energy bands, i.e., (1,10)eV, (0.1,100)GeV, (100,400)GeV, and (0.4,1.5)TeV, accommodating with the observed bands, are used to produce the modeling results. The brightness in sky maps is normalized to the same range, thus they share the same color bar. The horizontal white lines in the sky maps are the viewing angle of $\zeta=57^\circ$ for generating the modeling light curves in the middle panels. The observed optical light curve is adopted from \citet{slo09}. The GeV and sub-TeV light curves are adopted from Fermi \citep{abd13}, and MAGIC \citep{an16}.}
\label{CrabSM}
\end{figure*}

\subsection{The CR spectrum}

For the particle with a Lorentz factor $\gamma$, the individual CR spectrum at each position along the trajectory will be given by the following expression:
\begin{eqnarray}
F_{CR}(E_{\gamma},r)=\frac {\sqrt{3} e^2 \gamma} {2 \pi \hbar R_{CR}E_{\gamma} } F(x)\;,
\end{eqnarray}
where  $x=E_{\gamma}/E_{\rm cur}$, $E_{\gamma}$ is the energy of emitting photon, $E_{\rm cur}=\frac{3}{2}c\hbar \frac{\gamma^3}{R_{\rm CR}}$ is the characteristic energy of
the CR photon, $R_{\rm CR}$ is the local curvature radius at each point along the trajectory, and the function $F(x)$ is defined as
\begin{equation}
F(x)=x\int_{x}^{\infty}{K_{\rm 5/3}}(\xi)\;d\xi,
\end{equation}
where $K_{5/3}$ is the modified Bessel function of order $5/3$.

The total CR spectrum from all the radiation points will be the superposition of the individual particle spectra, which will be weighted by the surface primary particle flux as \citep{har08,har15}

\begin{equation}
\rm {\dot { N} }_p = \rm{ n_{GJ}c\pi{R^2_{\star}} \theta_{pc} \Delta A }\; ,
\end{equation}
where $\rm n_{GJ}$ is Goldreich-Julian particle density at the pulsar surface, $\rm R_{\star}$ the pulsar radius, $\rm \theta_{pc}$ the pulsar PC half-angle, and $\rm \Delta A $ the reduced surface area of the particle injection region relative to the PC area, which can be approximated by $\rm \Delta A= (r^{max}_{ovc})^2-(r^{min}_{ovc})^2 $.

\begin{figure}
\centering
\begin{tabular}{c}
\hspace*{-0.8cm}
\includegraphics[width=8.5 cm,height=6 cm]{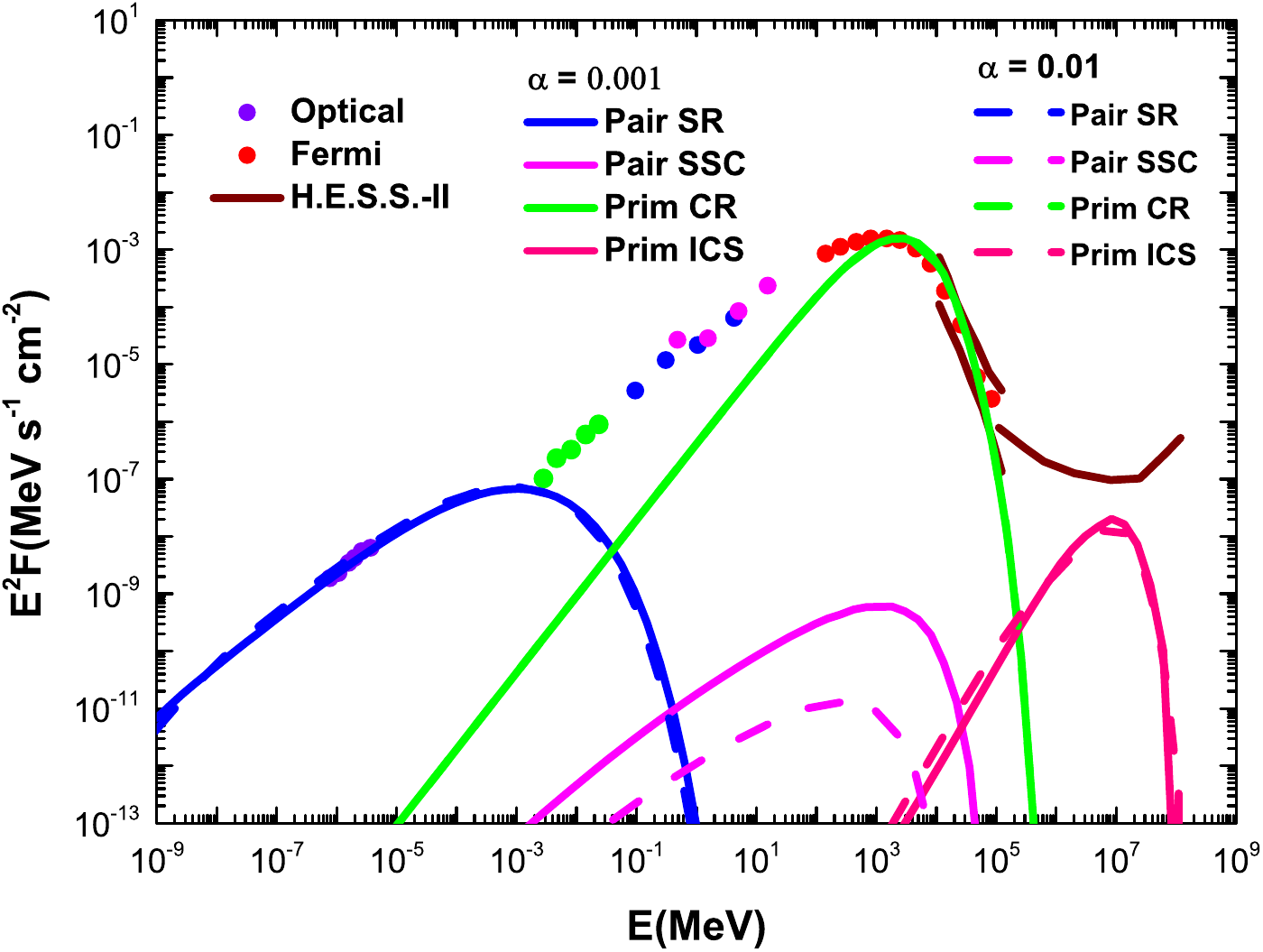}\,
\end{tabular}
\caption{The broadband SED of the Vela pulsar for the magnetic inclination $\chi=60^\circ$, viewing angle $\zeta=63^\circ$, $\kappa=3$, $\rm M_{pair}=2.4\times 10^4$ and $\rm \alpha=0.001$ shown as the solid lines. For comparison, the predicted SEDs for a larger pitch angle $\rm \alpha=0.01$ are also shown as the dashed lines. Observed data points are adopted from \citet{har02}, \citet{shi03}, and \citet{abd13}. The H.E.S.S.-II detection and 50 hr observational high-energy sensitivity are also plotted \citep{hol15, abda18}.}
\label{VelaSED}
\end{figure}

\begin{figure*}[htb]
\centering
\begin{tabular}{c}
\includegraphics[width=4.2cm,height=4.1cm]{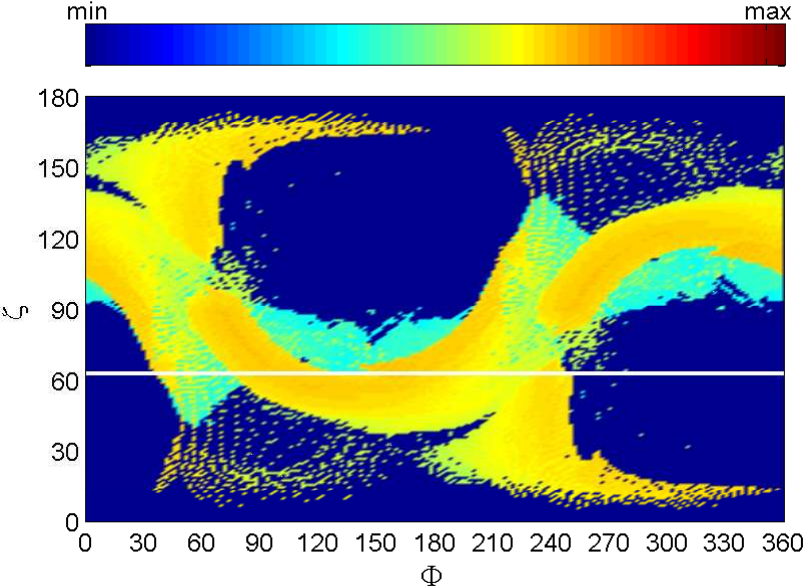}\,
\includegraphics[width=4.2cm,height=4cm]{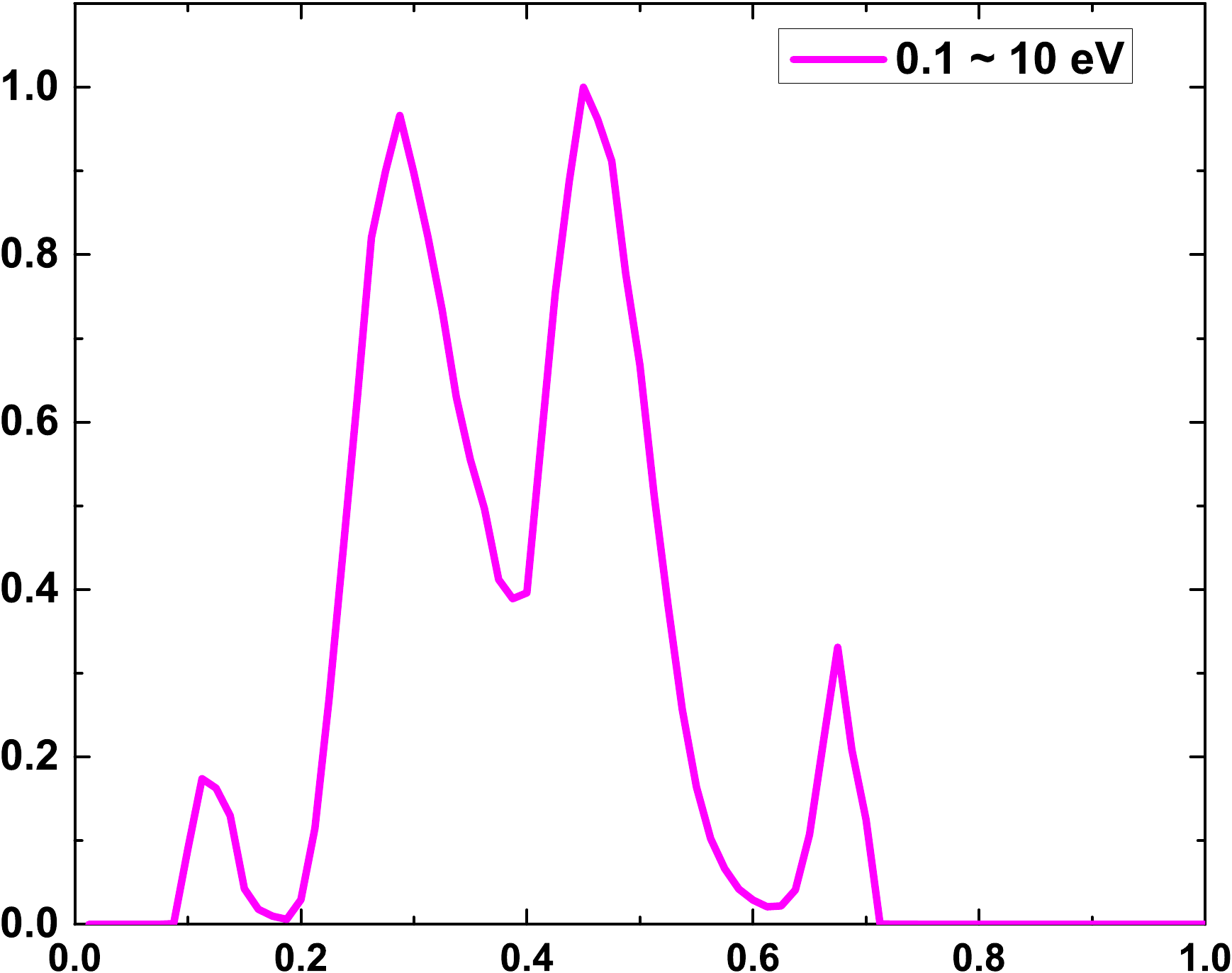}\,
\includegraphics[width=4.2cm,height=4cm]{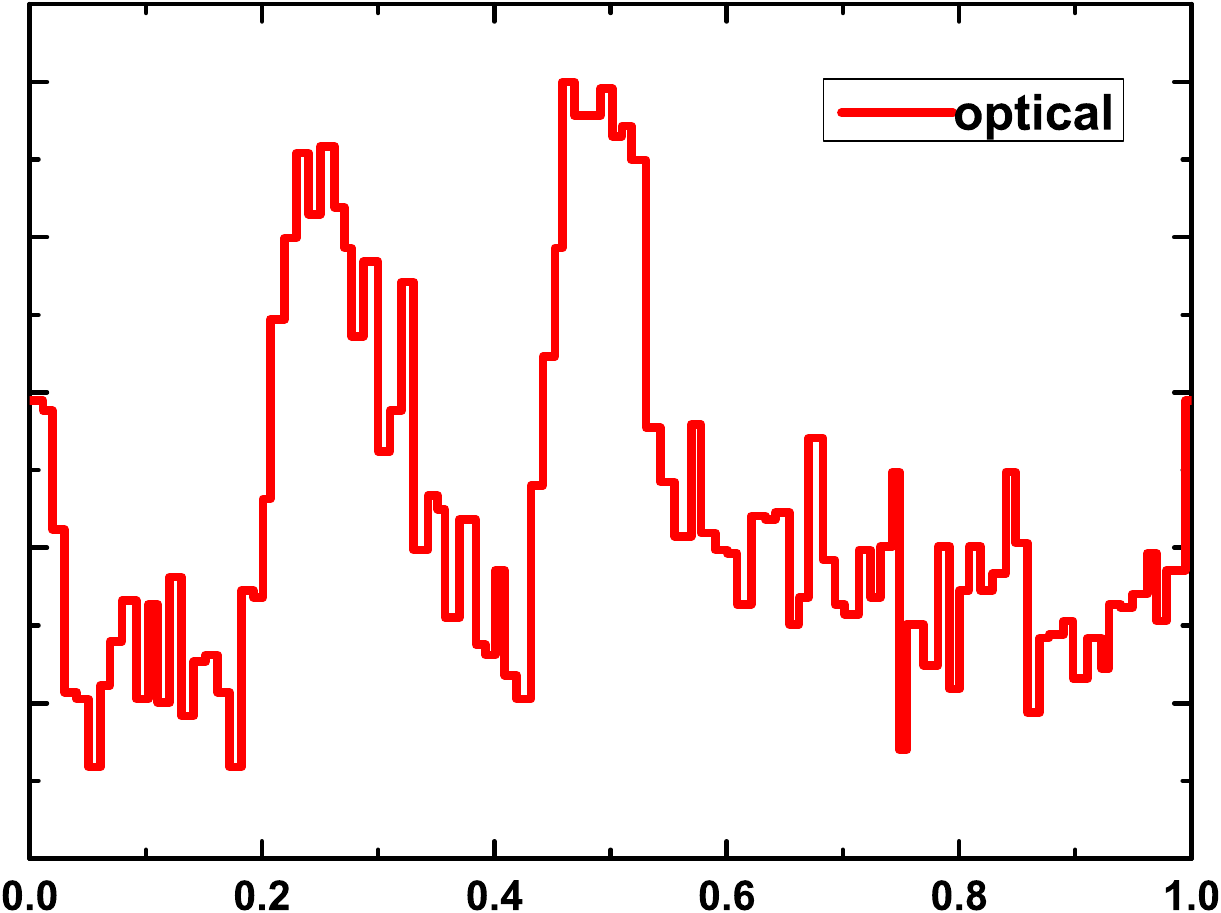}\\

\includegraphics[width=4.2cm,height=4.1cm]{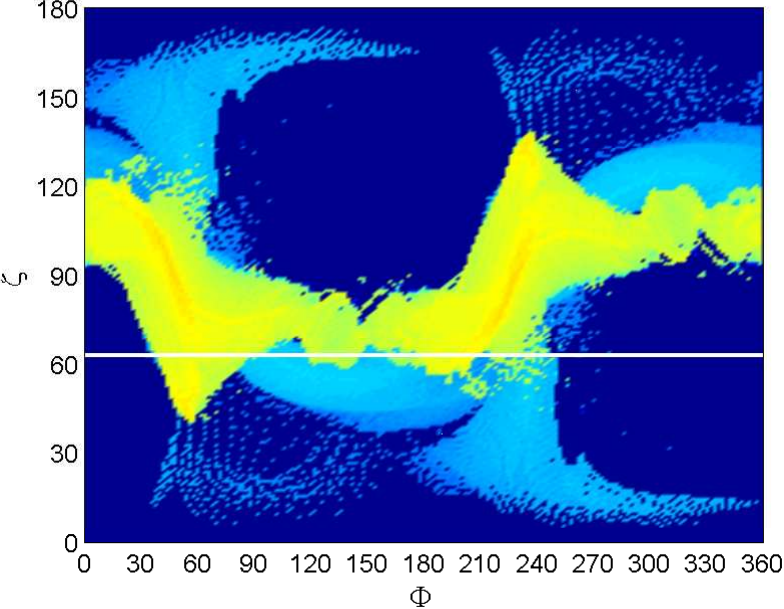}\,
\includegraphics[width=4.2cm,height=4cm]{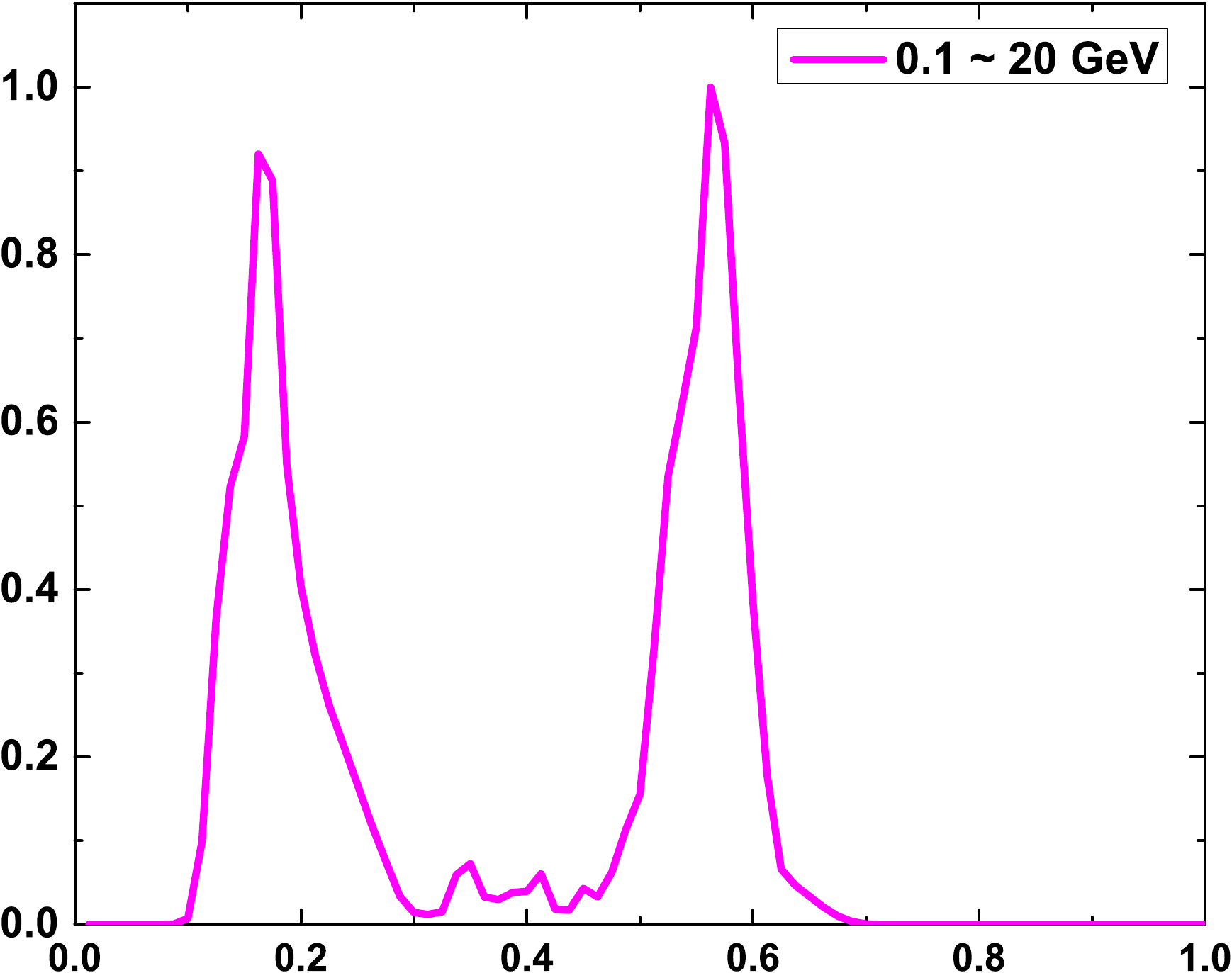}\,
\includegraphics[width=4.2cm,height=4cm]{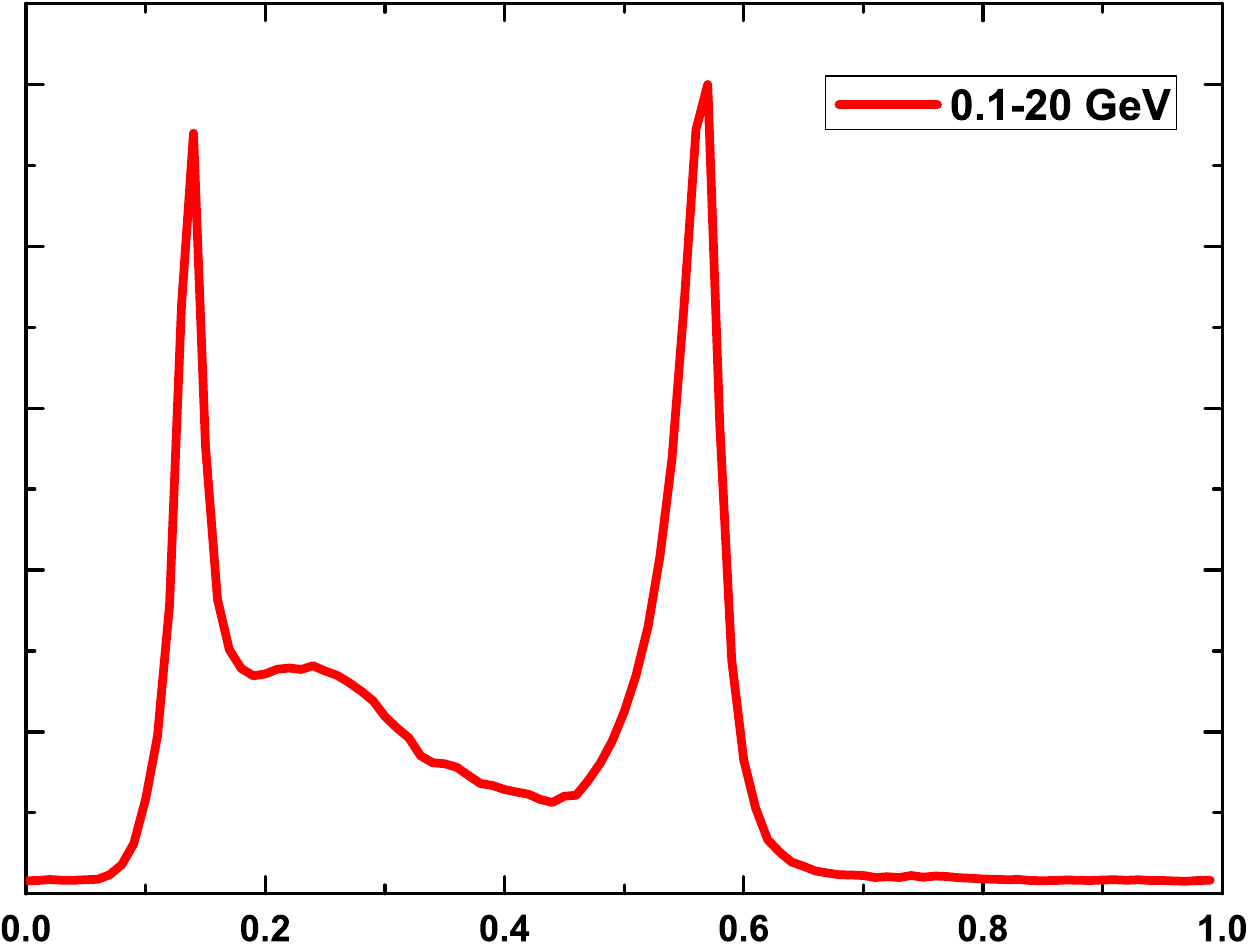}\\

\includegraphics[width=4.2cm,height=4.1cm]{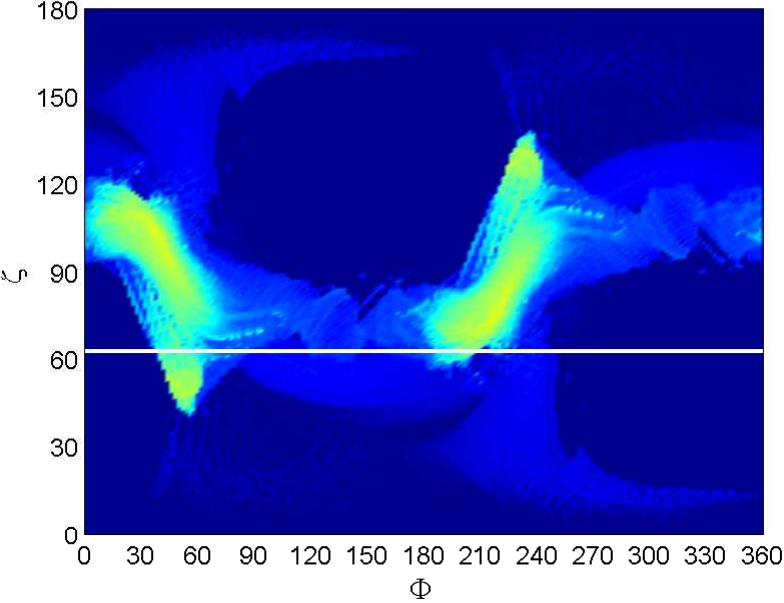}\,
\includegraphics[width=4.2cm,height=4cm]{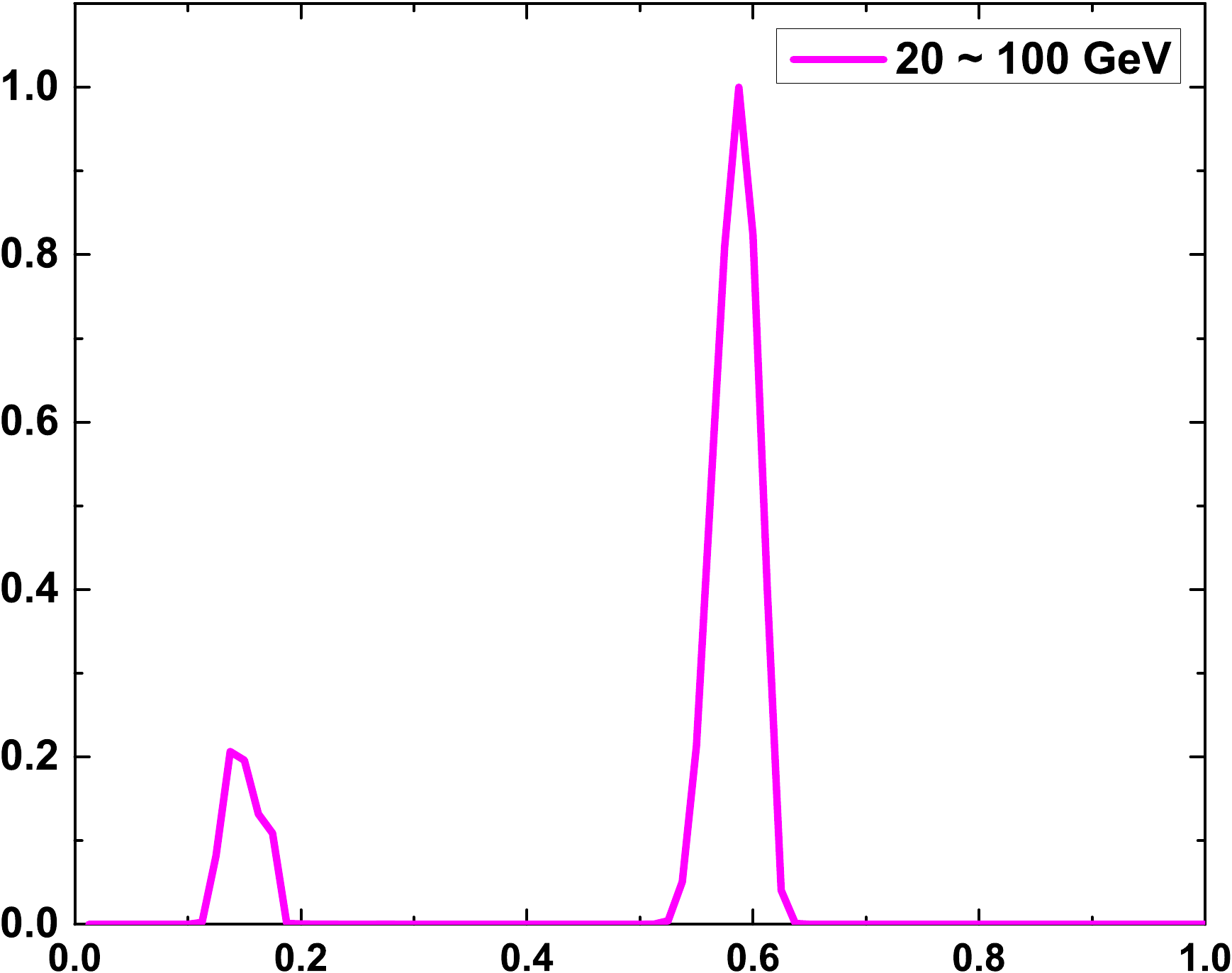}\,
\includegraphics[width=4.2cm,height=4cm]{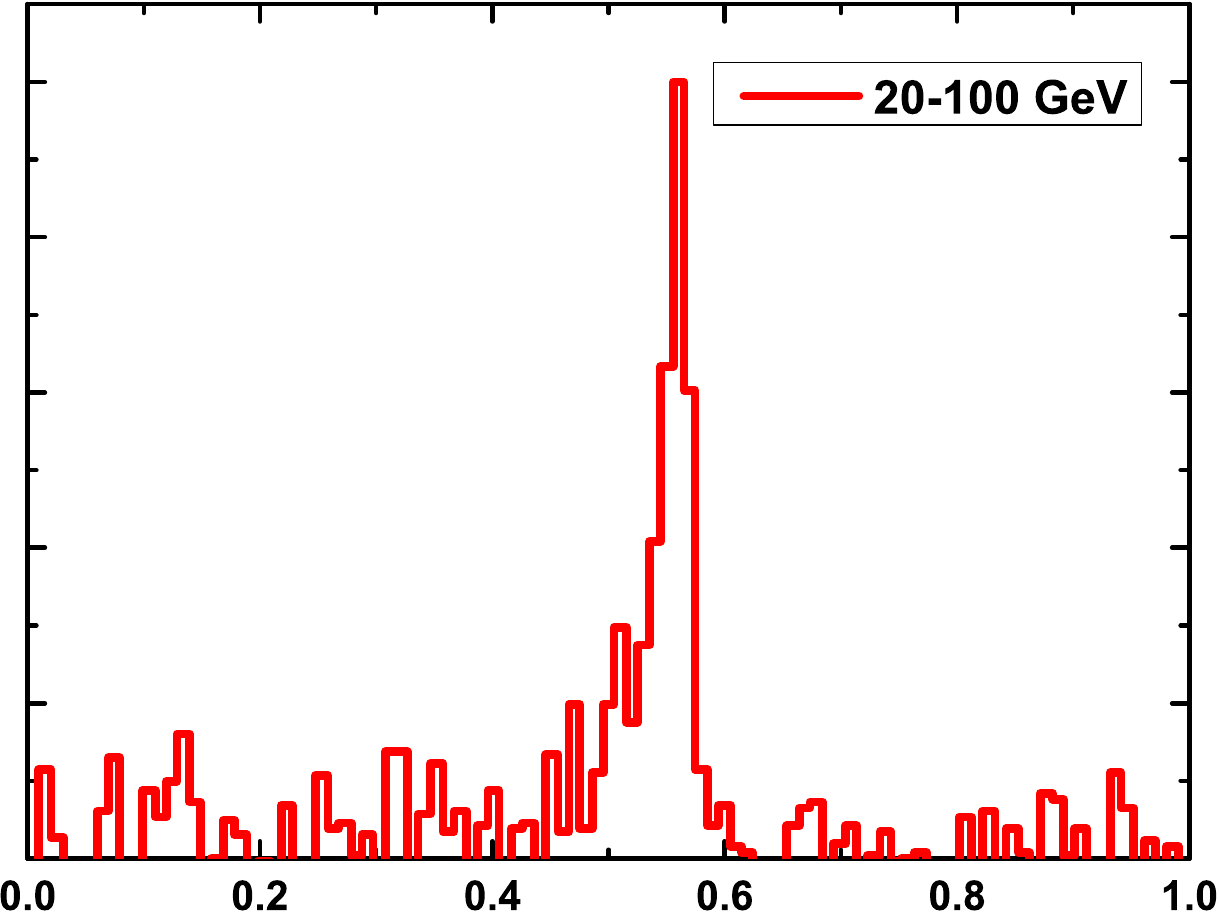}
\end{tabular}
\caption{The broadband sky maps and light curves for the Vela pulsar with the same parameters as in Figure \ref{VelaSED}. Three energy bands accommodated with the ones observed, i.e., (0.1,10)eV, (0.1,20)GeV, and (20,100)GeV, are used to produce the sky maps and the light curves. The horizontal white line in the left panels is the viewing angle of $\zeta=63^\circ$ for generating the counterpart light curves in the right panels. The observed light curves at those energies are taken from \citet{har00}, \citet{abd13} and \citet{abda18}.}
\label{VelaSM}
\end{figure*}

\subsection{The SR spectrum}\label{sub-SR}
The individual particle SR spectrum emitted by a particle with a Lorentz factor $\gamma$ at each emission point along the trajectory is given by the following formula:
\begin{eqnarray}
f_{SR}(E_{\gamma},r)=\frac {\sqrt{3} e^3 B \rm sin\alpha} {2 \pi \hbar m_e c^2 E_{\gamma} } F(x)\;,
\end{eqnarray}
where B is the magnetic field at each point along the trajectory, $\alpha$ is the pitch angle, $x=E_{\gamma}/E_{\rm SR}$, and $E_{\rm SR}= {\rm 3 \hbar \gamma^2 q_e B sin\alpha}/\rm 2 m_e c$ is the critical energy of an SR photon. The SR spectrum from all the secondary pairs will be the integration of the individual particle SR spectra with the normalized secondary particle distribution $\rm dN /{N_0 d\gamma}$, given by the following expression:
\begin{eqnarray}
F_{SR}(E_{\gamma},r)&=& \int_{\rm \gamma_{min}}^{\rm \gamma_{max}}f_{SR}(E_{\gamma},r)\left(\frac{dN}{N_0 d\gamma}\right)d{\gamma}.
\end{eqnarray}
The total SR spectrum of the secondary pairs should be weighted by the primary Goldreich-Julian particle density of the PC surface with the regulation of the pair multiplicity $\rm M_{pair}$ (the number of pairs produced by each primary particle), which is used to adjust the pair SR spectra to match the optical to hard-X-ray data and reads as $\rm {\dot {N} }_{pair} = M_{pair} {\dot { N} }_p$.

\subsection{The ICS/SSC spectrum}

In the paper, we compute the isotropic ICS/SSC of the primary particles and the secondary pairs scattering with the isotropic SR photons of the secondary pairs. The scattering spectrum of an individual particle with a Lorentz factor $\gamma$ at each emission point along the trajectory is given by
\begin{multline}
F_{ICS}(E_{\gamma},r)=\int_{\epsilon_1}^{\epsilon_2} \frac {2\pi r^2_0c}{\gamma^2}\frac{n(\epsilon)}{\epsilon}\\
\left[2q\ln q+(1+2q)(1-q)+\frac{(\Gamma q)^2(1-q)}{2(1+\Gamma q)} \right]d\epsilon \; .
\end{multline}
Here $n(\epsilon)=F_{SR}/ c r^2$ is the SR photon density, $q=E_1/\Gamma(1-E_1)$ with $E_1=E_{\gamma}/\gamma m_e c^2$ and $\Gamma=4 E_{\epsilon}\gamma m_e c^2$. The total primary particle ICS should be weighted by the primary particle flux $\dot{N}_p$. The total SSC spectrum from all the secondary pairs will be the integration of the individual particle SSC spectra with the normalized secondary particle distribution and weighted by the same particle flux as described in Section \ref{sub-SR}.

\subsection{The multiwavelength spectra and light curves}\label{sect-SM}
The location and geometry of the acceleration and radiation regions will be strongly imprinted on the pulsar multiwavelength radiation properties of the pulsar. Therefore, the pulsar multiwavelength light curve and spectra modelings of the pulsar provide a very powerful tool to diagnose the magnetospheric structure, the location of particle acceleration, and the emission mechanisms. In our model, all the particles are injected from the neutron surface -- the primary particles from the region of $\rm r_{ovc} = 0.9-1$ and for the secondary pairs from the adjacent region of $\rm r_{ovc} = 0.8-0.9$. Besides, the trajectories of both the primary particles and secondary pairs are separately tracked from the neutron surface up to $\rm 2.5 R_{LC}$. The secondary pairs cannot emit SR unless they acquire some pitch angles. While stringently constrained by the strong magnetic field, the particles will travel almost along it. \citet{ke15} discussed the evolution of the pitch angle in the strong magnetic field from analytical aspects. \citet{lyu98} and \citet{har08} proposed that the particles can acquire a small pitch angle at some height ($\rm r_{SR}$) due to the cyclotron resonance absorption. Therefore, we have choose the SR emission heights to be $\rm r_{SR} = 0.7 R_{LC}$, $\rm0.48 R_{LC}$, and $\rm 0.4 R_{LC}$ for Crab, Vela, and Geminga pulsars, respectively. We first compute the pair SR and collect the pair SR density along the particle trajectories from $\rm r_{SR}$ to $\rm 2.5 R_{LC}$. The pair SSC emissions are then computed by using the collected pair SR density from $\rm r_{SR}$ to $\rm 2.5 R_{LC}$. The primary particles cannot emit the CR and ICS unless their trajectories enter into the equatorial current sheet. Therefore, we track the trajectories of the primary particles from the neutron surface to up to $\rm 2.5 R_{LC}$ and only compute their CR and ICS when their trajectories enter into the equatorial current sheet, where the collected pair SR density is used for the target photons for ICS of the primary particles. The similar calculation methods are also used by \citet{har15} and \citet{har21}.

Assuming that the directions of the photon emission, $\bm{\eta}_{\rm em}$, are in the direction of the particle motion, $\bm{\beta}=\bm{v}/c$,  for the primary and secondary pairs, we can determine
the direction of the photon emission from each particle species using
\begin{eqnarray}
\mu_{\rm em}=\beta_{z}, \quad \phi_{\rm em}=\rm atan \left( \frac{\beta_y}{\beta_x} \right),
\end{eqnarray}
where  $\zeta=\rm acos(\mu_{\rm em})$ is the viewing angle.
The observed phase is obtained by adding the rotation and time-delay corrections:
\begin{eqnarray}
\Phi=\phi_{\rm rot}-\phi_{\rm em}-{\bf r_{\rm em}} \cdot {\bm \eta}_{\rm em}/R_{\rm LC},
\end{eqnarray}
where the first term, $\phi_{\rm rot}=\Omega \, dt$, is the rotation phase, the second term is the phase of the emitting photon, and the third term is the time-delay phase.

The multiwavelength sky maps can be produced by collecting all the emitting photons from each particle species in the ($\zeta$-$\phi$) plane. The corresponding light curves can be constructed by cutting the sky maps with a given viewing angle $\zeta$. More details about the constructions of the sky maps and the light curves can be found in those previous works \citep[e.g.,][]{kal14,har15,cao19,yang21}.

The  multiwavelength spectra can also be obtained by collecting all the radiation from both the primary particles and the secondary pairs at a given viewing angle along their trajectories using
\begin{multline}
F(E_{\gamma})=\frac{1}{4\pi d^2}\sum\limits_r [ F_{SR}(E_{\gamma},r)+F_{SSC}(E_{\gamma},r)+ \\
F_{CR}(E_{\gamma},r)+F_{ICS}(E_{\gamma},r) ] \;,
\end{multline}
where $d$ is the distance of the pulsar from the observer.

\begin{figure}[htb]
\centering
\begin{tabular}{c}
\hspace*{-0.8cm}
\includegraphics[width=8.5 cm,height=6 cm]{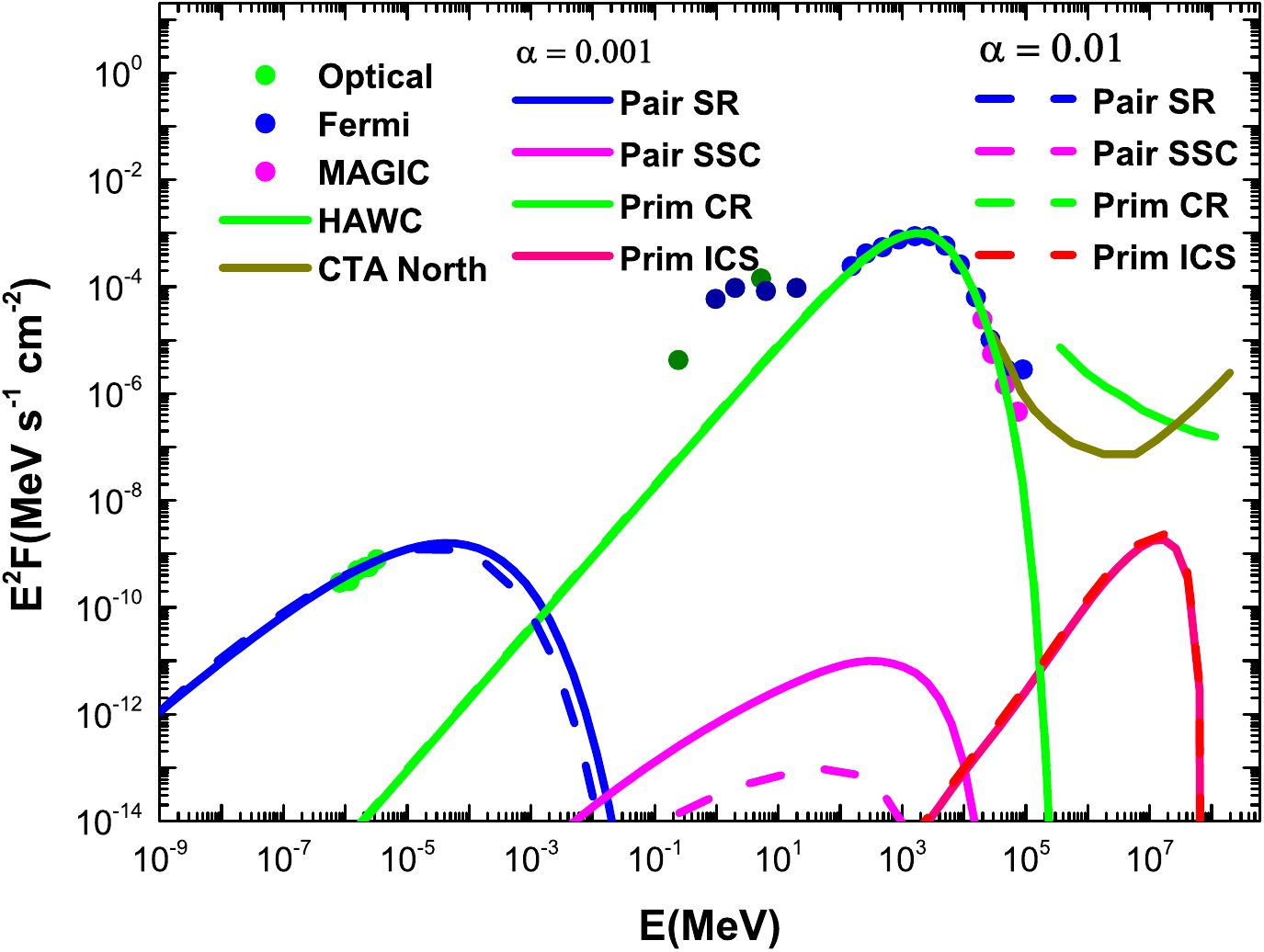}\,
\end{tabular}
\caption{The broadband SED of the Geminga pulsar for the magnetic inclination $\chi=70^\circ$, viewing angle $\zeta=66^\circ$, $\kappa=3$, $\rm M_{pair}=1\times 10^4$ and $\alpha=0.001$ shown as the solid lines. For comparison, the predicted SEDs for a larger pitch angle $\rm \alpha=0.01$ are also shown as the dashed lines. The observed data are taken from \citet{kui96},\citet{kar05}, \citet{shi06}, \citet{abd13}, and \citet{acc20}. The HAWC sensitivity curves \citep{abe17} and 50 hr CTA North observation curve \citep{ach18} are also plotted.}
\label{GemingaSED}
\end{figure}

\begin{figure*}[htb]
\centering
\begin{tabular}{c}
\vspace{0.3cm}
\includegraphics[width=4.2cm,height=4.1cm]{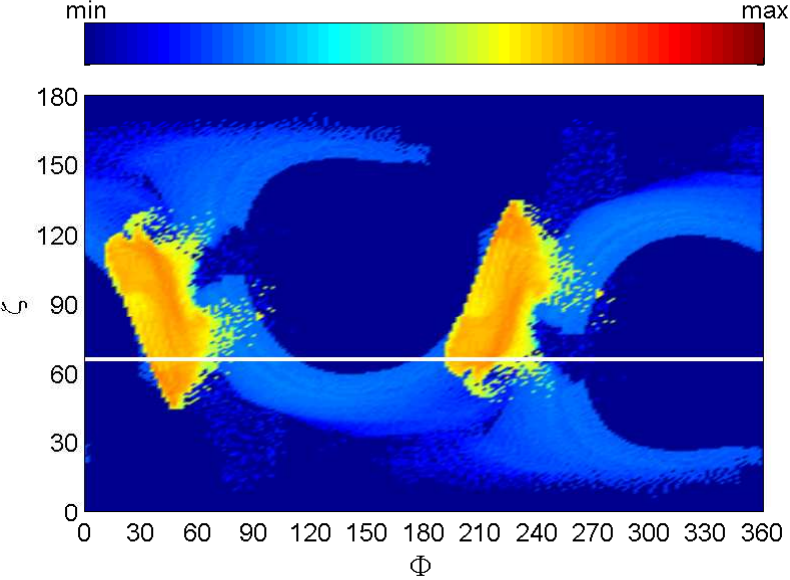}\,
\includegraphics[width=4.2cm,height=4cm]{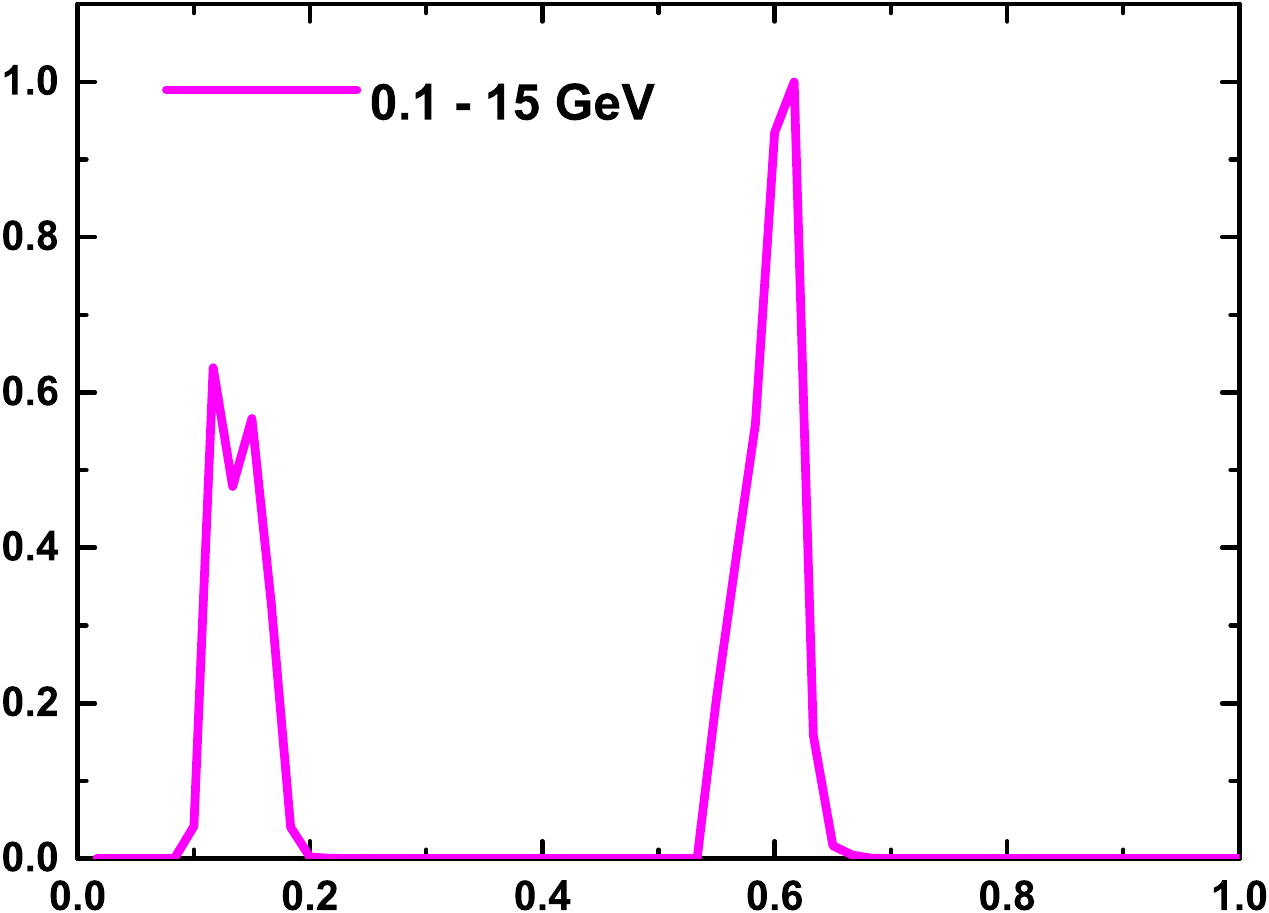}\,
\includegraphics[width=4.2cm,height=4cm]{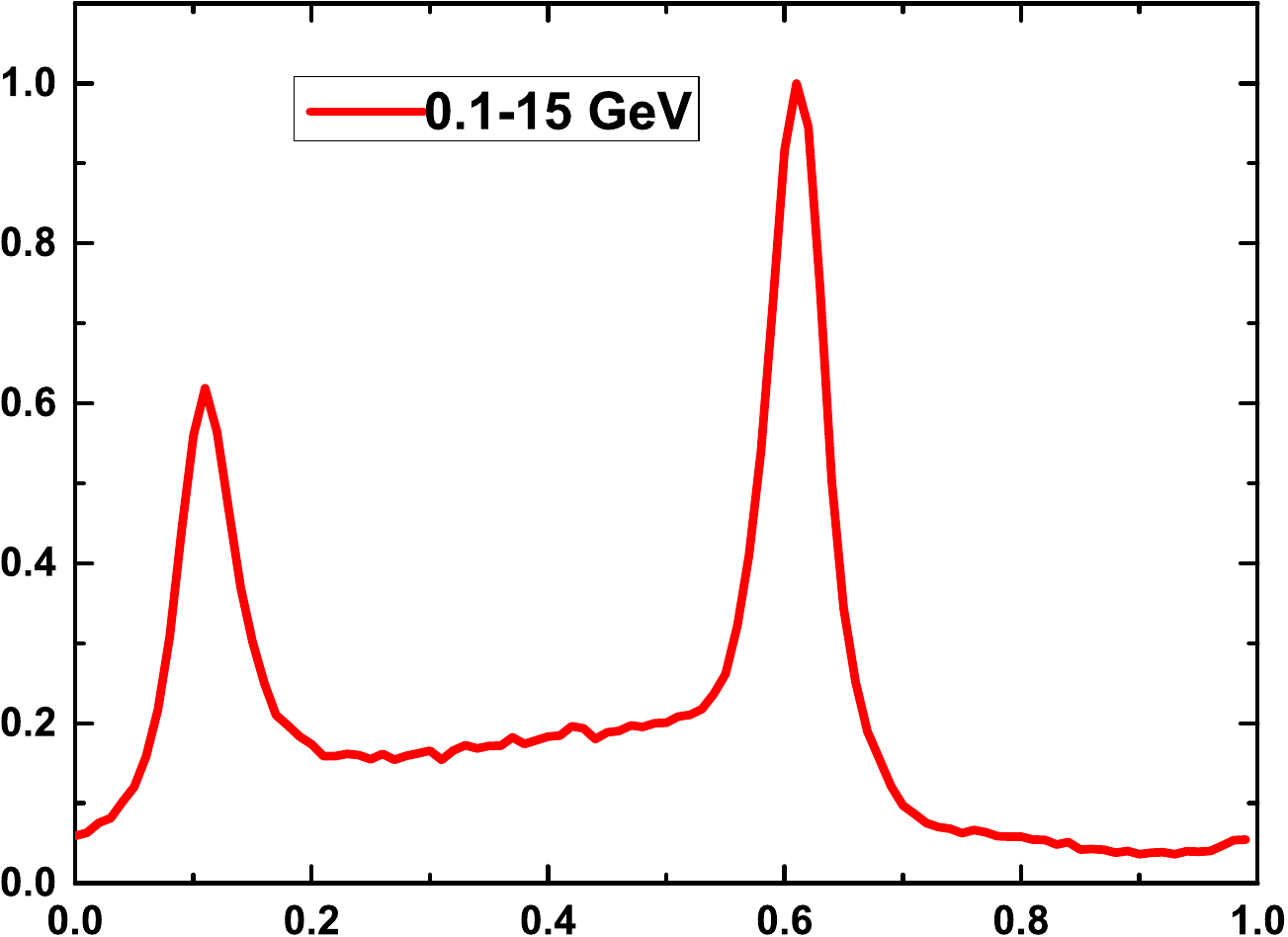}\\

\includegraphics[width=4.2cm,height=4.1cm]{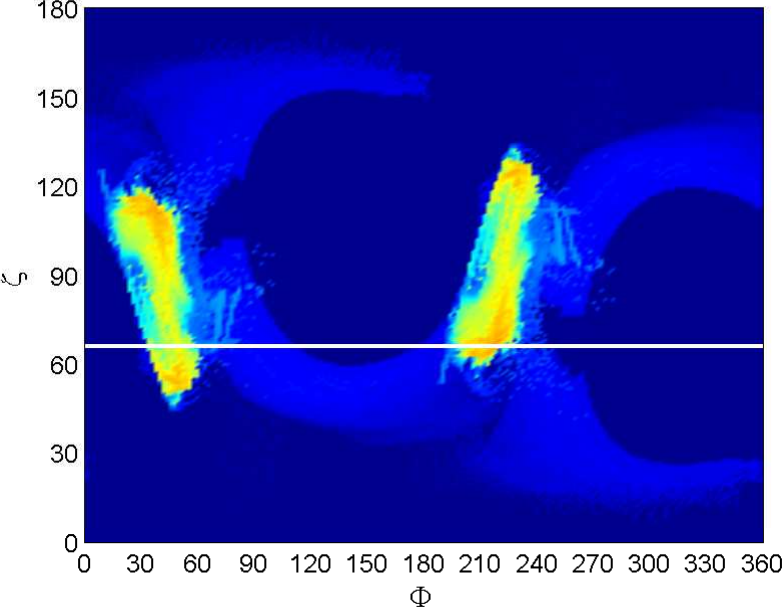}\,
\includegraphics[width=4.2cm,height=4cm]{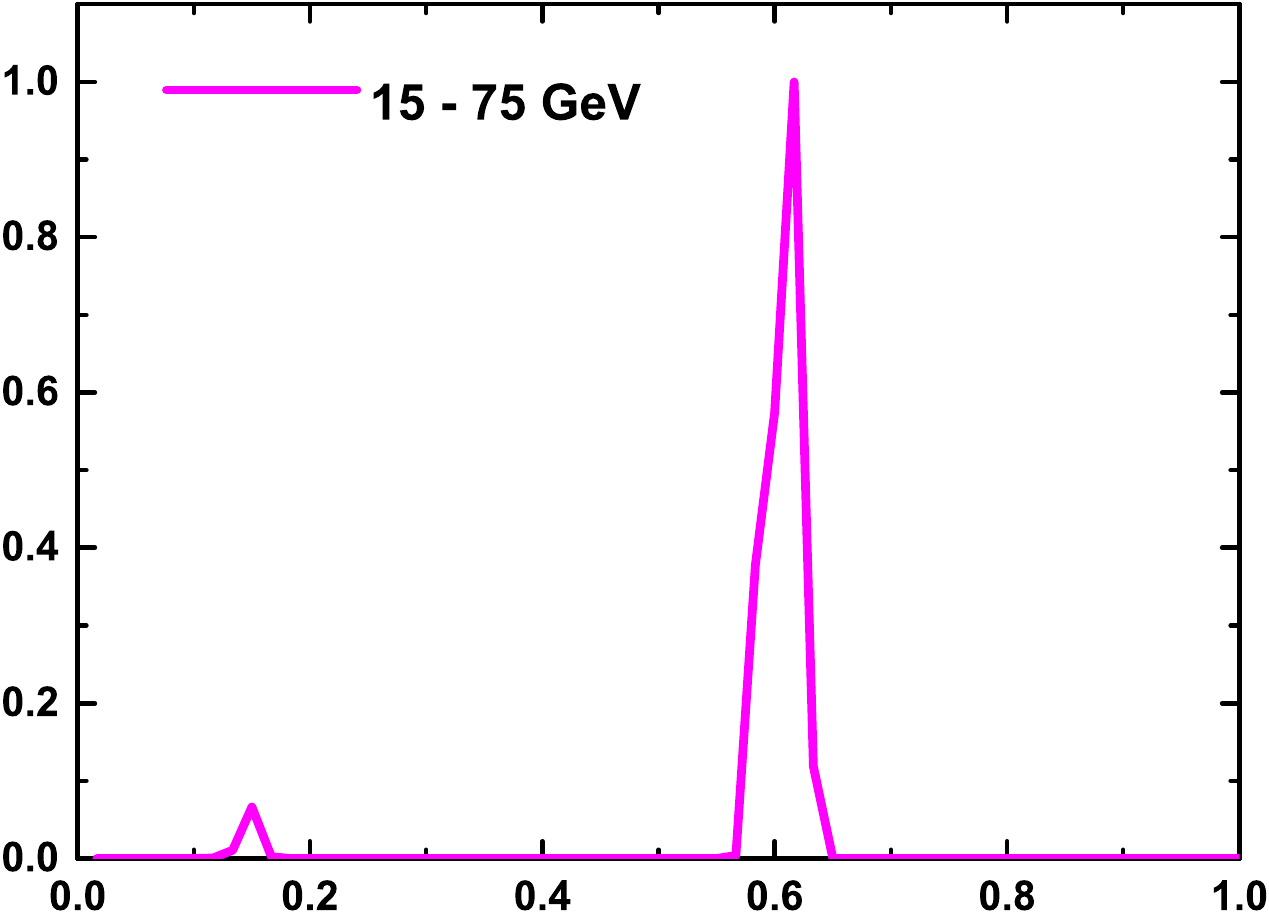}\,
\includegraphics[width=4.2cm,height=4cm]{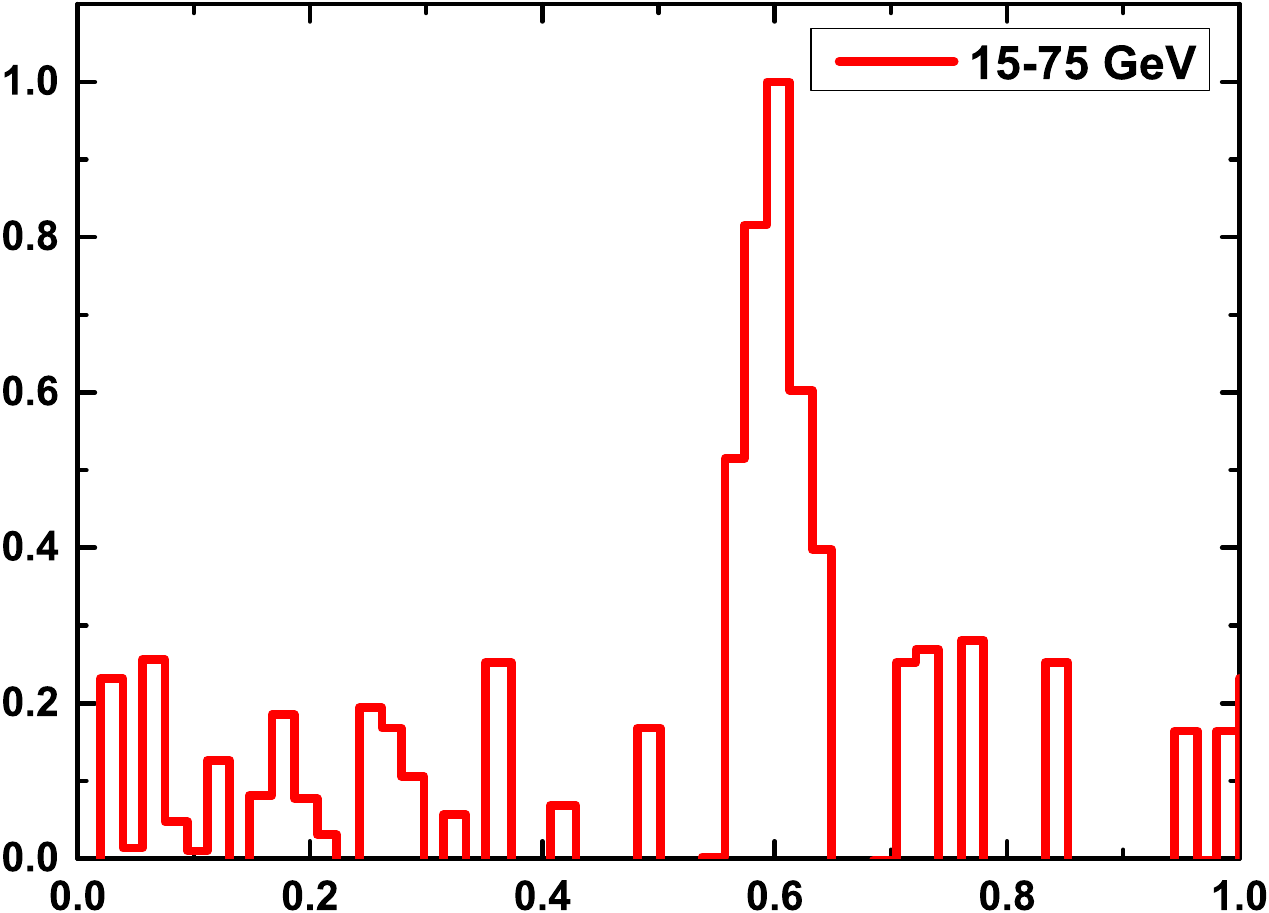}
\end{tabular}

\caption{The broadband sky maps and light curves for the Geminga pulsar with the same parameters as in Figure \ref{GemingaSED}. Two energy bands accommodated with the corresponding observed ones, i.e., (0.1,15)GeV and (15,75)GeV are used to generate the sky maps and the light curves. The horizontal white line in the sky maps is the viewing angle of $\zeta=66^\circ$ for generating the counterpart light curves. The observed light curves are taken from Fermi \citep{abd13} and MAGIC \citep{acc20}.}
\label{GemingaSM}
\end{figure*}

\section{The results}\label{Applications}
In the section, the dissipative model is used to compute the multiwavelength phase-averaged spectral energy distribution (SED), the sky maps, and the corresponding light curves for Crab, Vela, and Geminga pulsars. Two values of the pitch angle ($\alpha=0.01$ and $0.001$) are chosen to compare their effects on the SEDs in the following results. The pair cascade multiplicity $\rm M_{pair}$ is taken to be a free parameter to adjust the pair SR spectrum to match the optical to hard-X-ray data. Other parameters that affect the SED are magnetic inclination angle $\rm \chi$, the viewing angle $\rm \zeta$, and the secondary pair spectra.

\subsection{Crab pulsar}

The real observed parameters, surface magnetic field $\rm B=3.8\times 10^{12}~G$, period $\rm P=0.033~s$, and distance  $\rm d=2.0~kpc$ for the Crab pulsar, are used to obtain the modeling results. In Figure \ref{CrabSED}, we plot the broadband phase-averaged SED of the Crab pulsar for the magnetic inclination $\rm \chi=75^\circ$ and viewing angle $\rm \zeta=57^\circ$, with pitch angle $\rm \alpha=0.001$ and pair multiplicity $\rm M_{pair}=3\times 10^5$, shown as the solid lines in the figure. The viewing angle is chosen by the better match with the Fermi $\gamma$-ray data by CR of the primary particles. We also see that the the SR of the secondary pair can match well with the observed spectrum from the optical to the hard-X-ray band if the assumed pair spectra in Figure \ref{pairs} are used. The combination of the CR spectrum of the primary particles with the SSC spectrum of the pairs can well account for the transition from the GeV to the very-hight-energy spectrum up to around 1 TeV, observed by Fermi and MAGIC. Besides, the primary ICS is sensitive to the low-energy part of the pair SR, and scatters the SR photons in the Thompson limit. Therefore, the predicted ICS spectra from the primary particles produce a peak around 10~TeV that is close to the detection sensitivity of the High Altitude Water Cherenkov Observatory (HAWC).
For comparison, the predicted SEDs for a larger pitch angle $\rm \alpha=0.01$ are also shown as the dashed lines. When a larger $\rm \alpha$ is used, a lower pair $\rm \gamma_{cut}$ is required to reproduce the optical to the hard-X-ray spectra. Therefore, the predicted pair SSC spectra for a lower  $\rm \gamma_{cut}$ does not reach TeV energies to well match the MAGIC TeV spectra. Our results suggest that a low $\alpha$  $\rm \sim 10^{-3}$ is needed to reproduce the multiwavelength SEDs of the Crab pulsar. A similar low $\alpha$ is also used to explain the multiwavelength SEDs of the Crab pulsar by \citet{har21}. The magnetic inclination angle $\chi$ and viewing angle $\zeta$ are obtained by some previous the multiwavelength radiation modeling  with $\chi=60^\circ$ and $\zeta=60^\circ$ obtained by \citet{pet05}, $\chi=50^\circ$ and $\zeta=76^\circ$  by \citet{tang08}, $\chi=45^\circ$ and $\zeta=63^\circ$ by \citet{du12}, $\chi=75^\circ$ and $\zeta=63^{\circ}$ by \citet{bra15}, $\chi=75^{\circ}$ and $\zeta=50^{\circ}$ by \citet{yang21}, and $\rm \chi=45^{\circ}$ and $\rm \zeta=60^{\circ}$ by \citet{har15} and \citet{har21}. In fact, the inclination angles are not well constrained by observations and multiwavelength radiation modeling, but the viewing angle has been constrained from the measurement of the torus of the Crab wind nebula by \citet{ng08} with $\zeta \sim 61^\circ$. It is found that our modeling inclination angle and viewing angle are similar to those of the previous multiwavelength radiation modeling, and our modeling viewing angle of $\zeta=57^{\circ}$ is also close to that of the torus of the pulsar wind nebula with $\zeta \sim 61^\circ$ obtained by \citet{ng08}.

In Figure \ref{CrabSM}, we show the sky maps and the corresponding light curves from the optical to $\gamma$-ray bands. We observe that the intensities in the sky maps decrease with increasing energy, and that the bright patterns in the sky maps do not change significantly. This is the caustic effect of the particles whose trajectories can approach the equatorial current sheet and become nearly radial outside the LC. Besides, comparing the modeled light curves with the observed ones, it can be seen that we can systematically well reproduce the observed trends whereby the light curves display two narrow peaks that are approximately aligned in phase for different energy ranges. This arises from the result that the radiation generated by the primary and secondary particles comes from different but adjoint regions near the equatorial current sheet, which leads to the radiation arriving at almost the same phase. \citet{har21} also modeled multiwavelength light curves for the Crab pulsar by using the constant electric distribution in the current sheet of the FF magnetosphere. They found that a constant accelerating electric field distribution can not well reproduce the observed phases of multiwavelength light curves of the Crab pulsar. Our results show that a radially and azimuthally dependent accelerating electric field distributions from the simulations can provide an improved match to the multiwavelength light curves of the Crab pulsar.

\subsection{Vela pulsar}
The real observed parameters, surface magnetic field $\rm B=4\times 10^{12}~G$, period $\rm P=0.089~s$, and distance $\rm d=0.29~kpc$, are used to produce the multiwavelength light curves and spectra for the Vela pulsar.
In Figure \ref{VelaSED}, we plot the broadband SED for the Vela pulsar with magnetic inclination $\chi=60^\circ$ and viewing angle $\zeta=63^\circ$ with $\rm \alpha=0.001$ and $\rm M_{pair}=2.4\times 10^4$ as the solid lines. We see that the flux and the spectral index of the optical data can be well matched by the SR of the secondary pairs, and Fermi data are dominated by the CR of the primary particles accelerated in the equatorial current sheet. The high-energy tail of the primary CR spectra will approach or exceed the H.E.S.S.-II detection sensitivity, indicating that the H.E.S.S.-II has detected the high-energy tail of the primary CR component. However, our model cannot explain the observed emission in the MeV band, which may originate from primary SR radiation in the inner magnetosphere within the LC \citep{har15,har21}. Moreover, the primary ICS flux is very close to the H.E.S.S.-II detection sensitivity, therefore it is possible that the primary ICS component may partly explain observed TeV emission with the accumulated H.E.S.S.-II data in the future. For comparison, the predicted SEDs for a larger pitch angle $\rm \alpha=0.01$ are also shown as the dashed lines. We find that the pitch angle is not well constrained by observations due to the scarcity of the optical to hard-X-ray data. Therefore, a similar SED is also obtained by using a larger  $\rm \alpha$ value for the Vela pulsar. The magnetic inclination angle $\chi$ and viewing angle $\zeta$ are obtained by some previous multiwavelength radiation modeling, with $\chi=70^{\circ}$ and $\zeta=64^{\circ}$ obtained by \citet{du11}, $\chi=75^{\circ}$ and $\zeta=60^{\circ}$ by \citep{har15} and \citet{har21}, $\chi=60^{\circ}$ and $\zeta=50^{\circ}$ by \citet{bra15}, $\chi=70^{\circ}$ and $\zeta=79^{\circ}$ by \citet{ru17}, $\chi=75^{\circ}$ and $\zeta=65^{\circ}$ by \citet{har18}, $\chi=60^{\circ}$ and $\zeta=42^{\circ}$ by \citet{yang21}, $\chi=60^\circ$ and $\zeta=65^\circ$ by \citet{bar22}, and $\chi=65^\circ$ and $\zeta=64^\circ$ by \citet{cao24}. Similarly, the magnetic inclination is not well constrained by the current observation data and  multiwavelength radiation modeling, but the viewing angle has been constrained from the measurement of the torus of the Vela wind nebula by \citep{ng08}. We see that the modeled magnetic inclination and viewing angle are similar to those of the previous multiwavelength radiation modeling, and our modeled viewing angle $\zeta=63^{\circ}$ is also close to the observed one $\zeta=64^{\circ}$ obtained by \citet{ng08}.

In Figure \ref{VelaSM}, we show the sky maps and corresponding light curves from the optical to $\gamma$-ray bands. We can see that the bright caustics in the sky maps decrease with increasing energy, and that caustics produced by the secondary pairs distributed in more broader regions than those produced by the primary particles accelerated by the equatorial current sheet. This can be attributed to a significant inner emission from the secondary pair SR within the LC. Besides, the predicted optical light curves have two narrower peaks than the $\gamma$-ray one, which can well match with that observed. The predicted GeV light curves can well match the observed peak separation and relative ratio of the two peaks of the Fermi $\gamma$-ray light curves. The observed sub-TeV $\gamma$-ray light curves only visible in the second peak can also be well reproduced by our model. It is seen that our results can well reproduce the observed trends of the multiwavelength light curves for the Vela pulsar. Moreover, our results can well reproduce the decreasing ratio of the first $\gamma$ peak to the second one and the narrowing of the peak width toward the higher energies from the GeV to TeV bands. It is noted that we can give better match to the multiwavelength light curves of the Vela pulsar by including a radially and azimuthally dependent accelerating electric field distribution from the simulations than those obtained by \citet{har21}, where they used a uniform accelerating electric field independent of the radius and azimuth.

\subsection{Geminga pulsar}
The real Geminga observed parameters, surface magnetic field $\rm B=1.8\times 10^{12}~G$, period $\rm P=0.237~s$, and  distance $\rm d=0.25~kpc$ for, are used to produce the predicted multiwavelength light curves and spectra.
In Figure \ref{GemingaSED}, we plot the broadband SED for the Geminga pulsar with magnetic inclination $\chi=70^\circ$ and viewing angle $\zeta=66^\circ$ with $\alpha=0.001$ and $\rm M_{pair}=1\times 10^4$ in the solid lines. It can be seen that our results  well reproduce multiwavelength SEDs of the Geminga pulsar from the optical to $\gamma$-ray. We find that the optical data are produced by SR from the pairs, the Fermi data are dominated by the primary CR, and the MAGIC data are an extension of the primary CR. However, the predicted primary ICS spectra are below the current detection sensitivity. For comparison, the predicted SEDs for a larger pitch angle $\rm \alpha=0.01$ are also shown as the dashed lines. We find that the pitch angle is not well constrained by observations due to the scarcity of the optical to hard-X-ray data. Therefore, a similar SED is also obtained by using a larger  $\rm \alpha$ value for the Geminga pulsar. We note that the magnetic inclination $\chi$ and viewing angle $\zeta$ were obtained by some previous multiwavelength radiation modeling, with $\chi=50^\circ$ and $\zeta=86^\circ$ obtained by \citet{zhang01},  $\chi=60^\circ$ and $\zeta=90^\circ$ by \citet{pet09},  $\chi=45^\circ$ and $\zeta=87^\circ$ by \citet{bra15}, $\chi=75^\circ$ and $\zeta=55^\circ$ by \citet{har21}. It is found that the magnetic inclination angle and viewing angle were not well constrained; only the viewing angle is estimated as $\rm \zeta >60^{\circ}$ from the measurement of the Geminga X-ray image \citep{car03,pi15}. Although the magnetic inclination angles vary among different radiation models, our modeled viewing angle of $\zeta=66^{\circ}$ lies well in the range of $\rm \zeta >60^{\circ}$.

Figure \ref{GemingaSM} shows the sky maps and the corresponding light curves in the GeV to sub-TeV range. We see that the $\gamma$-ray sky maps show the characteristics of the caustic emission from the equatorial current sheet. The predicted GeV light curve is consistent with the one observed by Fermi, and the observed phases and the relative strength can be well reproduced by using the radially and azimuthally dependent accelerating electric field distributions from the simulations.
The predicted sub-TeV light curves are only visible in the second peak, and are in agreement with those observed by MAGIC.  Moreover, the peaks of the sub-TeV light curves are nearly aligned in phase with those of the GeV light curves, because they are both produced by the primary CR accelerated in the equatorial current sheet.
We also find that the ratio of the first $\gamma$-ray peak  to the second one decreases and the peak width also becomes narrow with the increasing energy, which can be reflected by the decreasing caustics in sky maps. It is seen that our results can well reproduce the observed trends of the multiwavelength light curves for the Geminga pulsar

\section{Conclusions and Discussions}\label{conclusions}
We have explored the properties of the pulsar multiwavelength light curves and spectra in the combined FF and AE magnetospheres. The combined FF and AE magnetospheres are computed by the pseudo-spectral method in the comoving frame, which can produce the self-consistent accelerating electric field distribution in the equatorial current sheet outside the LC. We use the AE velocity to define the trajectories of the primary particles and secondary pairs in the combined FF and AE magnetospheres. The pulsar multiwavelength light curves and spectra are computed by using multiple radiation mechanisms from both the primary particles accelerated by the accelerating electric field distribution in the equatorial current sheet and an assumed pair spectrum injected from the PC surface. Then, we perform a direct comparison between the predicted multiwavelength light curves and spectra and those of the Crab, Vela, and Geminga pulsars. Our results can well reproduce the observed trends of the multiwavelength light curves and spectra for the these three pulsars. Our results indicate that the high-energy tails of the primary CR spectra for the Vela and Geminga pulsars have been detected by H.E.S.S.-II and MAGIC. The primary ICS component above 1 TeV is detectable for the Crab and Vela pulsars by the current HAWC and future CTA. Moreover, our results can also systematically well reproduce the decreasing ratio of the first $\gamma$-ray peak to the second one toward higher energies, especially for the Vela and Geminga pulsars.

The pulsar multiwavelength light curves and spectra are also modeled by using a constant accelerating electric field distribution in the FF magnetosphere by \citet{har21}. However, they cannot well reproduce the peak phases of the light curves for those three pulsars. Our model can provide an improved match to the multiwavelength light curves of these three pulsars by using a radially and azimuthally dependent accelerating electric field distribution from the simulations themselves. However, our model cannot explain the bridge emission of the Fermi light curves and the $\sim$ MeV spectra. It is suggested that these emission components may originate in the inner magnetosphere within the LC \citep{har21,bar22}. We will explore the observed bridge emission and $\sim$ MeV spectra by introducing the accelerating electric field distribution not only in the equatorial current sheet outside the LC but also in the inner region within the LC in the future. Moreover, the polarization information can provide another powerful constraint on the locations of the emission regions and on the emission mechanisms. We will use the combined FF and AE magnetospheres to present the study of the pulsar polarization in the near future.

\acknowledgments
We would like to give our sincere gratitude to the anonymous referee for the valuable comments and suggestions that helped us improve the paper.
We acknowledge the support of the National Natural Science Foundation of China under the grants 12003026 and 12373045, and the Basic research Program of Yunnan Province under the grants 202001AU070070 and 202301AU070082.


\end{document}